\documentclass[aps,preprint,amssymb,amsmath,groupedaddress,superscriptaddress, longbibliography]{revtex4-1}

\usepackage[utf8]{inputenc}
\usepackage{graphicx}
\usepackage{epstopdf,epsfig}
\usepackage{amsmath}
\usepackage{amssymb}
\usepackage{amsbsy,amsxtra}
\usepackage{amsfonts}
\usepackage{braket}
\usepackage{dcolumn}   
\usepackage{bm}        
\usepackage{amssymb}   
\usepackage[usenames]{color}
\usepackage{natbib}
\usepackage{float}
\usepackage{hyperref}
\usepackage{subfigure}
\usepackage{lipsum}

\begin{document}
	

	\title{Tunnel Magnetoresistance in Self-Assemblies of Exchange Coupled Core/Shell Nanoparticles}

	\author{Fernando Fabris}
	\affiliation{Centro At\'omico Bariloche, CNEA-CONICET, Av. Bustillo
		9500,  Bariloche, R\'io Negro, Argentina}
	\author{Enio Lima Jr.}
	\affiliation{Centro At\'omico Bariloche, CNEA-CONICET, Av. Bustillo
		9500,  Bariloche, R\'io Negro, Argentina}
	\author{Cynthia Quinteros}
	\affiliation{Centro At\'omico Bariloche, CNEA-CONICET, Av. Bustillo
		9500,  Bariloche, R\'io Negro, Argentina}
	\author{Lucas Ne\~ner}
	\affiliation{Centro At\'omico Bariloche, CNEA-CONICET, Av. Bustillo
		9500,  Bariloche, R\'io Negro, Argentina}
	\author{Mara Granada}
	\affiliation{Centro At\'omico Bariloche, CNEA-CONICET, Av. Bustillo
		9500,  Bariloche, R\'io Negro, Argentina}
	\author{Mart\'in Sirena}
	\affiliation{Centro At\'omico Bariloche, CNEA-CONICET, Av. Bustillo
		9500,  Bariloche, R\'io Negro, Argentina}
	\author{Roberto D. Zysler}
	\affiliation{Centro At\'omico Bariloche, CNEA-CONICET, Av. Bustillo
		9500, Bariloche, R\'io Negro, Argentina}
	\author{Horacio E. Troiani}
	\affiliation{Centro At\'omico Bariloche, CNEA-CONICET, Av. Bustillo
		9500, Bariloche, R\'io Negro, Argentina}
	\author{V\'{\i}ctor Lebor\'an} \affiliation{Centro de Investigaci\'on en Qu\'imica Biol\'oxica e Materiais Moleculares (CIQUS), Dpto. de Qu\'imica-F\'isica, Universidad de Santiago de Compostela, 15782 Santiago de Compostela, Spain}
	\author{Francisco Rivadulla} \affiliation{Centro de Investigaci\'on en Qu\'imica Biol\'oxica e Materiais Moleculares (CIQUS), Dpto. de Qu\'imica-F\'isica, Universidad de Santiago de Compostela, 15782 Santiago de Compostela, Spain}
	\author{Elin L. Winkler}
	\email{winkler@cab.cnea.gov.ar}
	\affiliation{Centro At\'omico Bariloche, CNEA-CONICET, Av. Bustillo
		9500, Bariloche, R\'io Negro, Argentina}


	\begin{abstract}
		We report the precise control of tunneling magnetoresistance (TMR) in devices of self-assembled core/shell Fe$_3$O$_4$/Co$_{1-x}$Zn$_x$Fe$_2$O$_4$ nanoparticles ($0\leq x\leq 1$). Adjusting the magnetic anisotropy through the content of Co$^{2+}$ in the shell, provides an accurate tool to control the switching field between the bistable states of the TMR. In this way, different combinations of soft/hard and hard/soft core/shell configurations can be envisaged for optimizing devices with the required magnetotransport response.
		
	\end{abstract}
	
	\maketitle

\section{Introduction}
The possibility to manipulate the electrical resistive state of
magnetic/non-magnetic multilayers by an external magnetic field
(giant magnetoresistance, GMR) was demonstrated already 30 years
ago.\cite{Baibich, Binasch} The strong coupling between the electron
spin and charge degrees of freedom and the development of the tools
for their manipulation, triggered the growth of a new field called
spintronics.\cite{RevModPhys.76.323,Parkin2010} The fabrication of
magnetic tunnel junctions (MTJ) constitutes one of the most important
advances in this field since then.\cite{JULLIERE1975225,Moodera} A MTJ is composed of two layers of ferromagnetic conductors separated
by an insulating tunneling barrier, typically of $\approx$1 nm. The
different density of states at the Fermi level, N(E$_F$), of the spin
up/down subbands of the ferromagnetic metals imply a
spin-dependent tunneling probability. Therefore, the electrical
resistance of the device switches between high/low resistance states
 as the magnetic field changes the
relative orientation of the magnetizations of the two magnetic
layers (tunneling magnetoresistance, TMR). The MTJ devices  present high versatility and a great degree of functionalization, allowing to combine electrodes and barriers of different nature, where large tunneling magnetoresistance, up to hundreds of percents at room temperature, was obtained \cite{Yuasa2004,Djayaprawira2005}. However their fabrication is a challenge, involves advanced thin film deposition techniques and complex microfabrication procedures. Tunneling magnetoresistance has also been studied in simpler nanostructures as granular or disordered single films \cite{Lopez2003,Hwang1996}, where the grain boundaries act as tunnel junction barriers. However, the characteristic of the barrier cannot be controlled in these nanostructures and lower TMR values are obtained.

On the other hand the spectacular advances of the chemical synthetic methods produced over the last few years, offer an affordable route for the synthesis of complex nanostructures, with a precise control of their chemical composition, shape and size. \cite{Murray2000} 
These can be assembled in crystal-like structures over large
areas, in which the organic capping layer or a non-magnetic shell protecting the particles acts as a tunneling barrier that controls the electronic transport. \cite{Sun2000,Zeng2006,Black2000, Wang2009,Dugay2011, Kant2008} Spin-dependent  electrical transport and large magnetoresistance was also observed in devices formed by assembling of conducting magnetic nanoparticles (MNPs)\cite{Black2000,Taub2009,Tran2008} or binary nanoparticles superlattice \cite{Chen2013,Dong2010,Jiang2017}.
	
However, an important challenge  that must be addressed in this
field is the design of strategies to tune the switching field of the TMR devices, which is entirely determined by the anisotropy of the magnetic material.\cite{Lopez2003,Hwang1996,Kumar2013,Zhang2010,Helman1976,Elhilo1998} Therefore, a good handling over the coercivity of the magnetic nanoparticles would allow the control over the TMR of the assemblies, in a similar approach as that used in multilayers. \cite{Hu2002}

In this regard, an exciting  possibility is the fabrication of
devices based on self-assemblies of exchange coupled core/shell MNPs with tailored magnetic properties.\cite{LopezOrtega2015} The coercive field in these systems can be
finely modified through the interface magnetic
coupling\cite{Lavorato2017,Salazar-Alvarez2007,Skumryev2003,Sarkar2012, Winkler2012,Lavorato2016}, the
core size and shell thickness,\cite{Liu2015, Lottini2016,Lavorato2014} or the
magnetic anisotropy of the components.\cite{Sytnyk2013,Kumar2013,Fabris2019, Lavorato2015} Devices of this type should provide a way to manipulate at will the
characteristic switching field of TMR by controlling the magnetic
coupling across the core/shell interface. In this way, core-shell nanoparticles combine the properties of multilayered based tunnel junctions and granular or disordered thin films, offering very high versatility with a simple fabrication process.

Here we report the precise control of the TMR in self-assemblies of half metallic ferrimagnetic Fe$_3$O$_4$
nanoparticles encapsulated in ferrimagnetic electrical insulator
Co$_{1-x}$Zn$_x$Fe$_2$O$_4$ ($0\leq x\leq 1$). Progressive
replacement of Co$^{2+}$ by Zn$^{2+}$ in the shell reduces the
magnetic anisotropy and shifts the maximum of the TMR of the
self-assembled device in a perfect correlation with the magnetic response. These results demonstrate the feasibility of 
tunning the TMR switching field in self-assembled devices
formed by magnetic core/shell nanoparticles.

\section{Experimental Procedure}

Fe$_3$O$_4$/Co$_{1-x}$Zn$_x$Fe$_2$O$_4$  core/shell nanoparticles were fabricated by seed mediated high temperature decomposition of metal-acetylacetonates in benzyl ether assisted by oleic acid and
oleylamine, based on the method described in Refs. \cite{Sun2002, Sun2004,Fabris2019}. Initially monodisperse Fe$_3$O$_4$ seeds were obtained by mixing 12 mmol of Fe(III) acetylacetonate (Fe(acac)$_3$) with 24 mmol of 1,2-octanediol, 210
mmol benzyl ether, 8 mmol oleic acid, and 40 mmol oleylamine into a
three neck flask, under N$_2$ flow. The mixture was slowly heated up
to the reflux temperature (295 $^\circ$C) and  held for a total time
of 120 minutes. Then the solution containing the nanoparticles was
separated in five portions in order to overgrow the spinel ferrite
shell. At this point 0.6 mmol of
Co(II) acetylacetonate (Co(acac)$_2$) and Zn(II) acetylacetonate
(Zn(acac)$_2$) were added to the mixture, according to a nominal molar ratio
Co$_{1-x}$Zn$_x$Fe$_2$O$_4$, together with Fe(acac)$_{3}$ (1.2 mmol),
1,2-octanediol (3 mmol), oleic acid (3 mmol), oleylamine (3 mmol)
and benzyl ether (210 mmol), and the heating procedure was repeated.
Five samples with Zn nominal concentration x=0.00, 0.25, 0.50, 0.75 and
1.00 were synthesized. The samples were washed by adding ethanol
and centrifuged, followed by the addition of
acetone, and magnetically separated. Finally the MNPs were dispersed
in hexane. The nanoparticle composition were determined with an Inductively Coupled Optical Emission Spectrometer (ICP-OES) brand Agilent model 5110. To perform the measurements the samples were processed with a Berghoff microwave digester model SW4 in an acid mixture with $HNO_3:HCl$ $4:1$.

The self-assembly of the  core/shell MNPS was done at the liquid-air
interface following the procedure reported in  Refs. \cite{Dong2010,Chen2013, Jiang2017}. In the assembly process schematized in the upper panel of Figure \ref{Esquema}, a
drop of 10 $\mu$L of solution with 5 mg/mL of nanoparticles in hexane
is drop-casted onto the surface of triethylene glycol in a Teflon container, which was then covered by a glass slide.  In order to transfer the assemblies to a substrate, the teflon vessel of 1.5x1.5x1.0 cm$^3$ was designed with a 30$^\circ$ inclined base-plane where the substrate was located previously and was completed cover by the triethylene glycol. A self-assembled structure is formed after complete evaporation of hexane (between
10-15 min). After that the triethylene glycol was removed very slowly using a syringe in order to gently deposit the assembled film on the substrate. All the samples received a thermal treatment in a vacuum atmosphere ($\sim$10$^{-3}$ Torr) in order to reduce the organic coating of the particles and to promote a closer contact between them. The decomposition temperature of the organic nanoparticle coating was determined from thermogravimetric analysis. The self-organized nanoparticles were heated from room temperature up to 400 $^\circ$C at heating rate of 15$^\circ$C/min, kept at 400 $^\circ$C by 30 min and then cooled to room temperature at 15 $^\circ$C/min.

Structural characterization of core/shell powder samples was performed by conducting X-ray diffraction (XRD) experiments on a PANAlytical X$\textquoteright$Pert diffractometer with Cu K$\alpha$ radiation using a glass sample holder (step size 0.026$^o$, range 15$^o$-90$^o$). Transmission electron microscopy (TEM) images and electron diffraction patterns of powder samples and self assembled nanoparticles were taken in a Philips CM200 transmission electron microscope equipped with a Ultra-Twin lens operating at 200 kV and a resolution of 0.19 nm. In order to perform the structural characterization of the self-assemblies of core/shell nanoparticles, they were transferred from the triethylene glycol surface to commercial silicon nitride TEM grids followed by thermal annealing. Atomic force microscopy (AFM) measurements were done in a Veeco (now Bruker) Dimension 3100 SPM in tapping mode using a standard tip. The 2 $\mu$m scans were done using a scan frequency of 1Hz and after waiting 30 minutes for thermal stabilization and noise reduction. No modification of the surface was observed after the measurements.

The magnetic properties were studied using a commercial superconducting quantum interference device magnetometer (SQUID, MPMS Quantum Design). To perform the measurements the self-assembled nanoparticles were  transferred them from the triethylene glycol surface to glass substrate  (4 mm $\times$ 6 mm) followed by the thermal annealing. The magnetoresistive devices were fabricated by thermal evaporation of the Au/Cr electrodes on glass substrates. The Au/Cr patterns of 7 $\mu$m channel length and 6 mm channel width, were fabricated by photolithography as shown in the middle panel of Figure \ref{Esquema}. Then, the self-assembled of core/shell nanoparticles floating on the triethylene glycol surface were transferred to the prepatterned glass substrates, and the obtained films were thermally annealed. The magnetotransport measurements were performed using a Keithley 4200 source-measure unit in a two probe configuration, with a maximum applied field of $\pm$12 kOe.

\section{Results and discussion}
\begin{figure*}
	\includegraphics[width=0.9\textwidth]{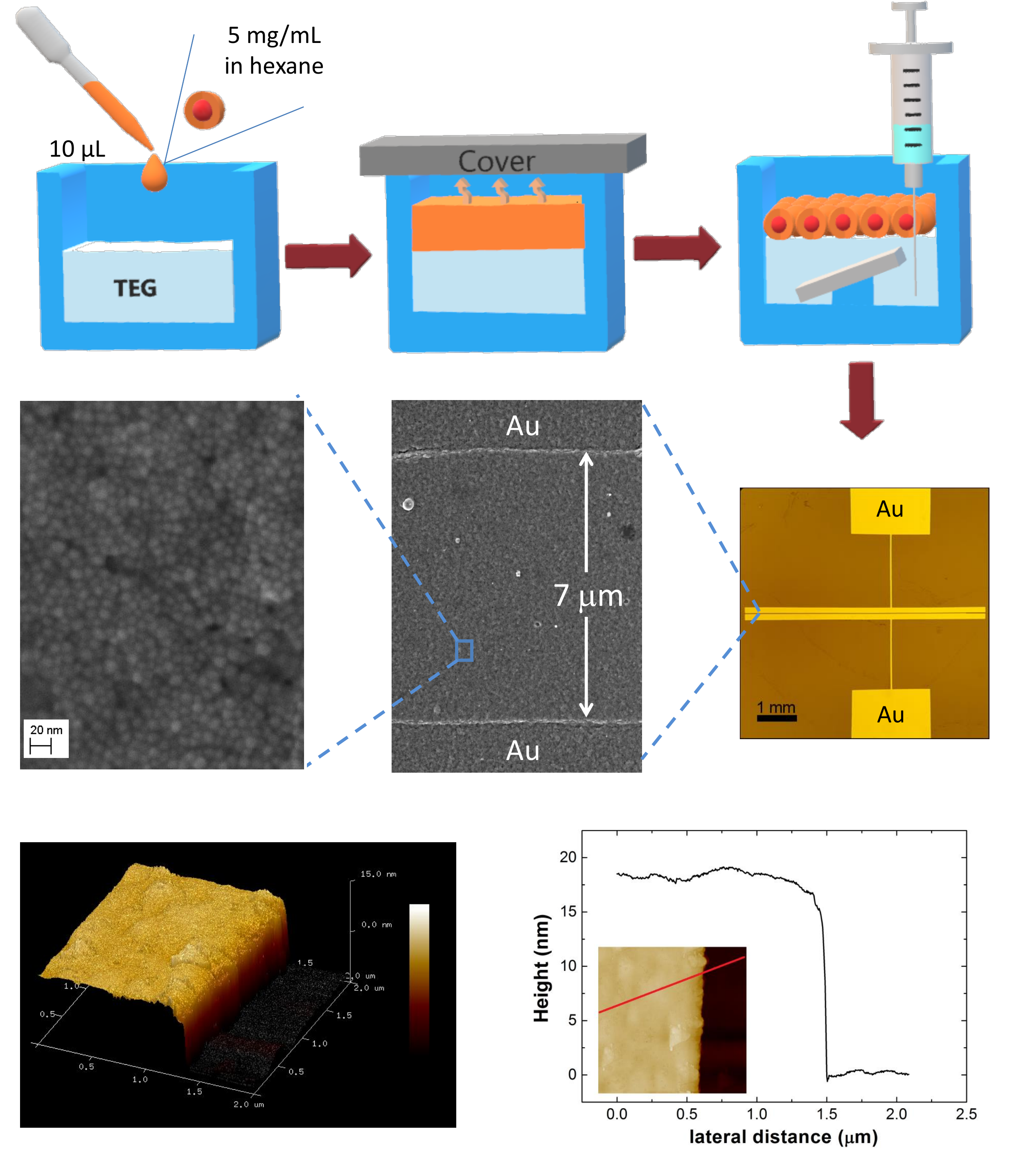}
	\caption{Scheme of the process of self-assembly of the particles at
		liquid-air interface and the transfer of the assemblies on a
		substrate. The middle part of the figure shows an optical image (right), along with two SEM images of the glass substrate patterned with Au
		electrodes, separated by 7 $\mu m$, with the self-assembled nanoparticles
		deposited on top. The bottom part of the figure shows the topography and height profile cross section of the annealed nanoparticles assemblies analyzed by atomic force microscopy (AFM). The height profile across the film edge shows a thickness of $\approx$ 20 nm, consistent with a two-layer assembly.} 
	\label{Esquema}
\end{figure*}

\begin{figure*}\centering
	\includegraphics[width=0.9\textwidth]{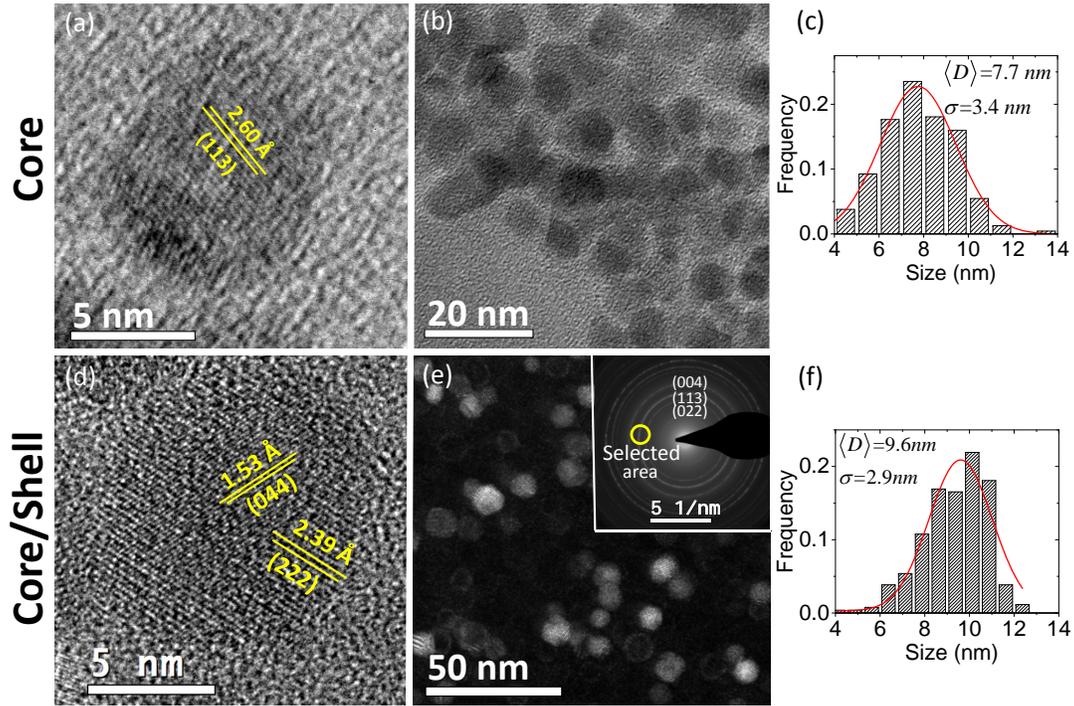}
	\caption{Top panels: (a-b) TEM images of the Fe$_3$O$_4$ core nanoparticles annealed at 400 $^\circ$ in vac uum atmosphere, HRTEM where the interplanar distance of (113) spinel phase are signaled, and (c) size histogram fitted with a Gaussian function. Bottom panels:(d) HRTEM image of Fe$_3$O$_4$/CoFe$_2$O$_4$ nanoparticles annealed at 400 $^\circ$ in vaccuum atmosphere, where different crystalline orientations (044) and (222) of spinel structure for core and shell are indicated. (e) Dark field TEM image reconstructed with a fraction of the (113) of the spinel diffraction ring and (f) size dispersion for the core/shell nanoparticles with the corresponding Gaussian fitting.}
	\label{TEM_core}
\end{figure*}

Figure \ref{TEM_core} compares the morphology and size, measured by TEM, of Fe$_3$O$_4$ nanoparticles seeds with a representative core/shell system, Fe$_3$O$_4$/CoFe$_2$O$_4$, both samples were subjected to the same thermal annealing at 400 $^\circ$C in vacuum atmosphere. From the size histograms, fitted with a Gaussian function, the mean particle sizes ($\langle D\rangle$) were calculated, resulting 7.7 nm and 9.6 nm, for core and core/shell systems, respectively. From high resolution TEM (HRTEM) image it is noticed that the core is monocrystalline and the shell growth epitaxial over the core for most of the nanoparticles. Moreover different crystalline orientations for core and shell can be observed for most of the nanoparticles as noticed in Figure \ref{TEM_core}(d) where the (044) and (222) crystalline planes of spinel phase are signaled for core and shell, respectively. The core/shell structure is confirmed by dark field, as shown in Figure \ref{TEM_core}(e) where the TEM image was recorded with a small objective aperture positioned on the (113) brighter electron diffraction ring of the spinel phase. In this way the bright contrast in the reconstructed image corresponds to the spinel grains with the selected crystallographic orientation as schematically drawing in  Figure S1(d-e) (see Suplemental Material). From the HRTEM and dark field TEM images, and from the comparison with the core size, the thickness of the Co$_{1-x}$Zn$_x$Fe$_2$O$_4$ shell was estimated as $\sim$ 1 nm. The composition of the nanoparticles was analyzed by inductively coupled plasma optical emission spectrometry (ICP-OES). From this analytical technique the concentration of the transition metal ions for all the samples were obtained and reported in Table \ref{ICP}. From these data, and assuming a Fe$_3$O$_4$ core of 7.7 nm of diameter, we calculated the shell stoichiometry which shows a systematic evolution consistent with the nominal concentration.

\begin{table}
	\caption{Total concentration of transition metal ions in core/shell nanoparticles measured from inductively coupled plasma mass spectrometry (ICP-OES). The last column presents the calculated shell stoichiometry considering the mean  nanoparticles size obtained by TEM and a Fe$_3$O$_4$ core of 7.7 nm of diameter.}
	\label{ICP}
	\begin{tabular}{l|ccc|c}
		\hline
		Samples & \multicolumn{3}{c|}{Total concentration}  & Calculated \\
		&\%Co & \%Zn&\%Fe & shell stoichiometry\\
		\hline
		$x=0.00$ & 9.40   & 0.00  & 90.60 & Co$_{0.61}$Fe$_{2.39}$O$_4$\\
		$x=0.25$ & 7.71   & 2.36  & 89.93 & Co$_{0.45}$Zn$_{0.15}$Fe$_{2.40}$O$_4$ \\
		$x=0.50$ & 6.01   & 4.80  & 89.19 & Co$_{0.39}$Zn$_{0.35}$Fe$_{2.26}$O$_4$ \\
		$x=0.75$ & 2.68   & 5.78  & 91.54 & Co$_{0.18}$Zn$_{0.44}$Fe$_{2.38}$O$_4$ \\
		$x=1.00$ & 0.00   & 9.94  & 90.06 & Zn$_{0.80}$Fe$_{2.20}$O$_4$\\
		\hline
	\end{tabular}
\end{table}
	\begin{table}[ht]
		\caption{Nanoparticles size distribution parameters of Fe$_3$O$_4$ and Fe$_3$O$_4$/Co$_{1-x}$Zn$_x$Fe$_2$O$_4$ obtained from the TEM images, where $\langle D \rangle$ is the mean particle size, and $\sigma$ is the standard deviation. The last columns present the mean blocking temperature  $\langle T_B \rangle $ and the effective magnetic anisotropy constant $K_{eff}$ calculated from the ZFC-FC magnetization curves.}
		\label{NPsize}
		\begin{tabular}{lcccc}
			\hline
			Sample  & $\langle D\rangle$  & $\sigma$  & $\langle T_B \rangle $  & $K_{eff}$ \\
			&nm &nm &K&10$^{6}$erg/cm$^{3}$\\
			\hline
			Fe$_{3}$O$_{4}$ & 6.9 & 2.6 &  17 & 0.32 \\
			$x=0.00$ & 9.6  & 2.9 & 191 & 1.54 \\
			$x=0.25$ & 9.9  & 3.8 & 171 & 1.25 \\
			$x=0.50$ & 9.5  & 2.9 & 135 & 1.12 \\
			$x=0.75$ & 9.4  & 2.7 & 115 & 0.99 \\
			$x=1.00$ & 9.2  & 3.3 &   8 & 0.07 \\
				\hline
		\end{tabular}
	\end{table}

The self assemblies of Fe$_3$O$_4$-core/Co$_{1-x}$Zn$_x$Fe$_2$O$_4$-shell nanoparticles were obtained by the liquid-air interface process \cite{Dong2010,Chen2013, Jiang2017} as explained in the Experimental Procedure Section.   In order to perform the different measurements, the self-assembly and the subsequent thermal treatment was reproduced using a commercial silicon nitride support grid for TEM characterization, and a glass substrate patterned with two Au electrodes separated by $\sim$7 $\mu$m (as shown in the SEM image of Figure \ref{Esquema}) for the magnetotransport studies. The topography of the annealed assemblies was analyzed by atomic force microscopy. Images acquired at different regions of the films reveal a large homogeneity with uniform and smooth surface, as observed in the bottom panel of Figure \ref{Esquema}. From the AFM height profile cross section at the film boundary, an average film thickness of 20 nm was measured, which corresponds to two layers of nanoparticles.

Homogeneity and narrow size distribution are essential conditions to reach large area of self-organization;  for core/shell nanoparticles, as observed from Figure \ref{TEM2} and Figure S1(a) (see Supplemental Material \cite{SupMat}), assemblies of several microns are obtained. From  Figure S1(f) it is also noticed that the self-organized nanoparticles are separated by a gap of $\sim$ 1 nm. As the thermogravimetric analysis indicates that approximately 7\% of residual mass remains in the systems after the thermal treatment at 400 $^\circ$C in vacuum atmosphere, and infrared spectroscopy measurements do not detect  organic molecules (see Figure S3 in Supplemental Material \cite{SupMat}), we conclude that the gap between the nanoparticles is formed by amorphous carbon. The nanoparticles size distributions measured from the TEM micrographs, are shown in the middle panels of Figure \ref{TEM2}. From the fitting of the histograms with a Gaussian function, the mean nanoparticles size  $\langle D\rangle$ was calculated and summarized in Table \ref{NPsize}, which vary between 9.2-9.9 nm for all the systems.  We also notice that the nanoparticle size and also the superstructure of the self-assembly is preserved at higher annealing temperature, however at 600 $^\circ$C the nanoparticles start to coalesce (see Figure S2 Supplemental Material \cite{SupMat}). From the HRTEM images (shown in Figure \ref{TEM_core} for $x=$0 and in Figure \ref{TEM2} for $x=$0.75 and  $x=$1) it is observed that the core/shell microstructure is preserved after the annealing at 400 $^\circ$C, where  different interplanar distances and crystallographic planes orientation for the inner and outer part of the particle can be measured. As mentioned before, this morphology is confirmed by dark field images, as shown in Figure \ref{TEM2} for x=0.00, 0.25 and 0.50. Although the core/shell microstructure is preserved, we can not discard some degree of interdiffusion at the interface as reported for similar nanoparticles systems \cite{Lopez2012,Krycka2013}; however, as we are going to discuss later, the magnetoresistance measurements confirms the half-metallic nature of the Fe$_{3}$O$_{4}$ core.

\begin{figure*}[!]
	\centering
	\includegraphics[width=0.7\textwidth]{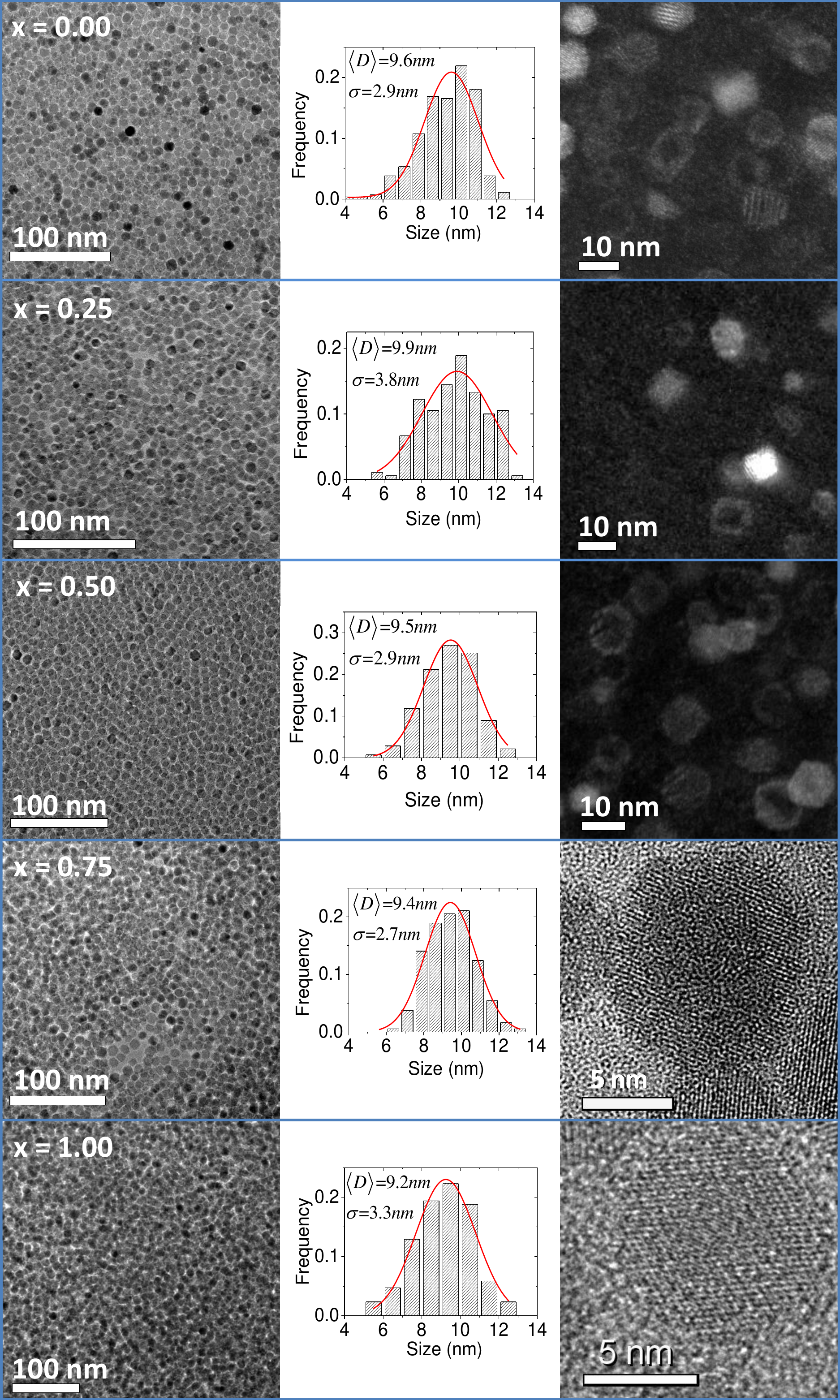}
	\caption{TEM images of the self-assembled
		Fe$_3$O$_4$/Co$_{1-x}$Zn$_x$Fe$_2$O$_4$ core/shell nanoparticles for
		the different compositions from $x=0$ to $x=1$, along with the
		corresponding size distribution histograms. The right panel show
		representative images where the core/shell structure can be
		appreciated from dark field images for $x=0$, $0.25$ and $0.50$ and from high resolution TEM for $x=0.75$ and $1$. The bright contrast in the dark field images corresponds to the spinel grains with the particular crystallographic orientation selected by  positioning the small objective TEM aperture on the brighter (113) electron diffraction ring of the spinel phase. }
	\label{TEM2}
\end{figure*}

\begin{figure}[ht]
        \includegraphics[width=0.5\textwidth]{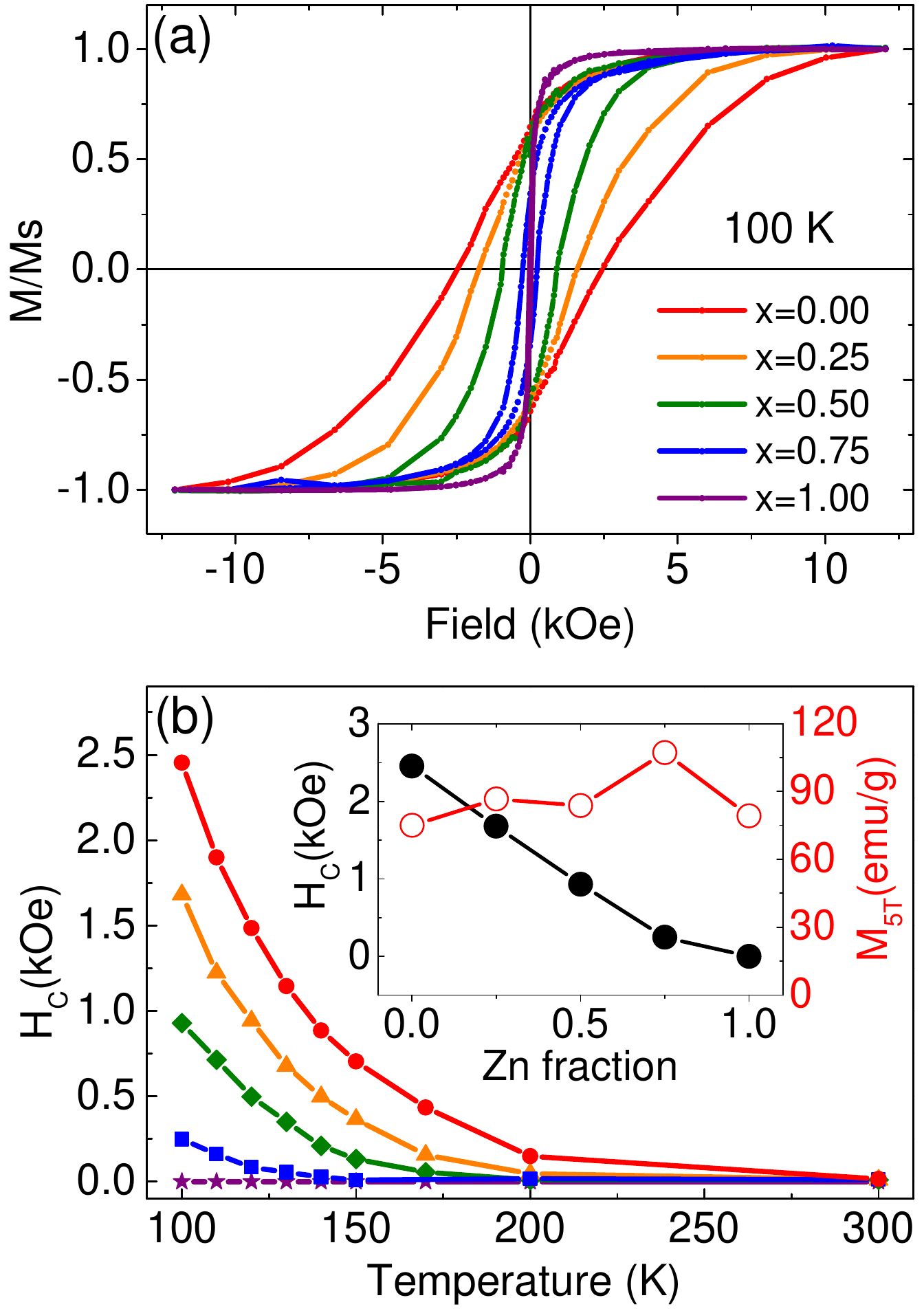}
\caption{(a) Hysteresis loops measured at 100 K, and (b) coercive field as a function of the temperature for the annealed	Fe$_3$O$_4$/Co$_{1-x}$Zn$_x$Fe$_2$O$_4$ core/shell nanoparticles assemblies. The inset shows the H$_{C}$ for the annealed self-assembled particles as a function of $x$, and also the magnetization measured at 5 T (M$_{5T}$) for powder sample annealed under the same conditions as the self-assemblies. Both measurements were acquired at 100 K.}
    \label{HC}
\end{figure}

Figure \ref{HC}(a) and Figure S4 (see Supplemental Material \cite{SupMat}) show the magnetization hysteresis loops of the annealed self-assembled nanoparticles, measured with the magnetic field applied in the plane of the substrate. The hysteresis loops show a single magnetization reversal for all compositions, with the coercive field ($H_C$) decreasing approximately linearly with the Zn concentration $x$, as shown in Figure \ref{HC}(b). Notice that the magnetization hysteresis loop reported for binary nanocrystal superlattice CoFe$_2$O$_4$$-$Fe$_3$O$_4$ results from the superposition of the individual component and an annealing treatment is an essential step to obtain an exchange interaction between the half-metallic Fe$_3$O$_4$ and the magnetic insulator CoFe2O4 \cite{Chen2013}. On the contrary, in the core/shell morphology, both phases are strongly coupled at the interface and homogeneous loop is obtained even in the as-synthesized system as shown in Figure S5. The magnetization inversion process of Fe$_3$O$_4$/Co$_{1-x}$Zn$_x$Fe$_2$O$_4$ core/shell nanoparticles can be analyzed from the theoretical approach developed for bimagnetic soft/hard exchange coupled nanostructures.\cite{Skomsky1993, Kneller1991,Zhao2006,Zhao2005,Leineweber1997} From these studies a critical size for the soft magnetic component, $\delta_{crit}$, was found bellow which  both phases are rigidly coupled by interface exchange interaction and reverse their magnetization in a coherent mode at the nucleation field, $H_N$. This critical size is approximately twice the magnetic Bloch wall width of the hard phase $\delta_{w}=\pi\sqrt{A/K}$, where $A$ is the exchange stiffness. In this regime, a single square hysteresis loop is obtained, with $H_C=H_N=2(K_{c}f_{c}+K_{sh}f_{sh})/(M_{c}f_{c}+M_{sh}f_{sh})$; where $M$, $f$ and $K$ are the magnetization, film thickness (or volume fraction), and magnetic anisotropy of the core ($c$) and shell ($sh$), respectively.\cite{Fullerton1999,LopezOrtega2015,Zhao2007b} Instead, if the size of the soft magnetic phase is larger than $\delta_{crit}$ exchange spring behavior is found, the magnetization reversal is nonuniform and lower coercivities are obtained. \cite{Fullerton1999,Zhao2005,Zhao2006}  For CoFe$_2$O$_4$ the values reported for $\delta_{w}$ span in the range of  13-20 nm,\cite{Coey2010, LopezOrtega2015} larger than the diameter of the Fe$_3$O$_4$ soft core used in this work. Therefore, rigid exchange coupling of the magnetizations of the core/shell phases is expected. This conclusion is supported by the study of Fe$_3$O$_4$/CoFe$_2$O$_4$ soft/hard bilayers, where a critical thickness of 8 nm was found for the crossover from rigid coupling to exchange spring behavior as a function of the Fe$_3$O$_4$ thickness\cite{Lavorato2016b}.  Therefore, the decrease of  $H_C$ with $x$ can be accounted by the diminution of the magnetic anisotropy of the shell from $\sim$4 10$^{6}$erg$/$cm$^{3}$ for $x=0$, to $\sim$2 10$^{5}$erg$/$cm$^{3}$ for $x=1$ \cite{Bullita2014} in agreement with the expression for $H_C$ in the rigid exchange coupled regime.

The systematic dependence of the magnetic anisotropy with the shell composition  is also reflected in the zero-field-cooled (ZFC) and field-cooled (FC) magnetization  curves shown in Figure S6 of Supplemental Material \cite{SupMat}. The temperature where the change from blocked to the superparamagnetic regime is observed decreases progressively with increasing $x$. From the
maximum of the energy barrier distribution calculated as
$f(T_{B})=(1/T)(d(M_{ZFC}-M_{FC})/dT)$ the mean blocking
temperature, $\langle T_{B}\rangle$, and the effective magnetic
anisotropy constant, $K_{eff}=27$ $k_{B}T_{B}/v$, can be obtained
for particles of total volume $v$, as reported in Table \ref{NPsize}.\cite{Dormann1996} For comparison, Table S1 (see Supplemental Material \cite{SupMat}) reports the  parameters that characterize the magnetic properties of dispersed nanoparticles before the annealing process. Both systems, the dispersed nanoparticles and the annealed assemblies,  present qualitative and quantitative similar behavior, with an enhancement of the effective magnetic anisotropy when the concentration of Co in the shell increases, which points out that the magnetic behavior is governed by the	hard/soft rigid coupling magnetization inversion process, as analyzed previously. However, the annealed assemblies present an approximately 20\% larger $\langle T_{B}\rangle$, $H_C$ and  $K_{eff}$, probably due to an increase of the dipolar interaction and the improvement of the crystallinity in the annealed assemblies.  


The magnetic measurements demonstrate that the effective magnetic anisotropy of the system can be controlled by adjusting the shell composition, without appreciably modifying the morphology and the overall magnetic saturation, as observed from the inset of Figure \ref{HC}(b). Given that the anisotropy of the system is to a great extent responsible of the switching field of the TMR, devices
made of self-assembled core/shell nanoparticles provide an ideal system for studying spin-dependent transport between magnetic nanoparticles. Although both materials at the core/shell structure are strongly exchange coupled and behave as a unique magnetic entity with an average magnetic anisotropy, the conductivity of each phase is different. While the Fe$_3$O$_4$ core is half-metallic, the Co$_{1-x}$Zn$_{x}$Fe$_{2}$O$_{4}$ shell is a semiconductor. Therefore, in order to observe TMR properties it is crucial to have Fe$_3$O$_4$ phase to provide the spin polarized transport, whereas the role of the shell is to modulate the switching field by tuning the magnetic anisotropy, while acting as a tunnel barrier.

\begin{figure}[h!]
	\includegraphics[width=0.5\textwidth]{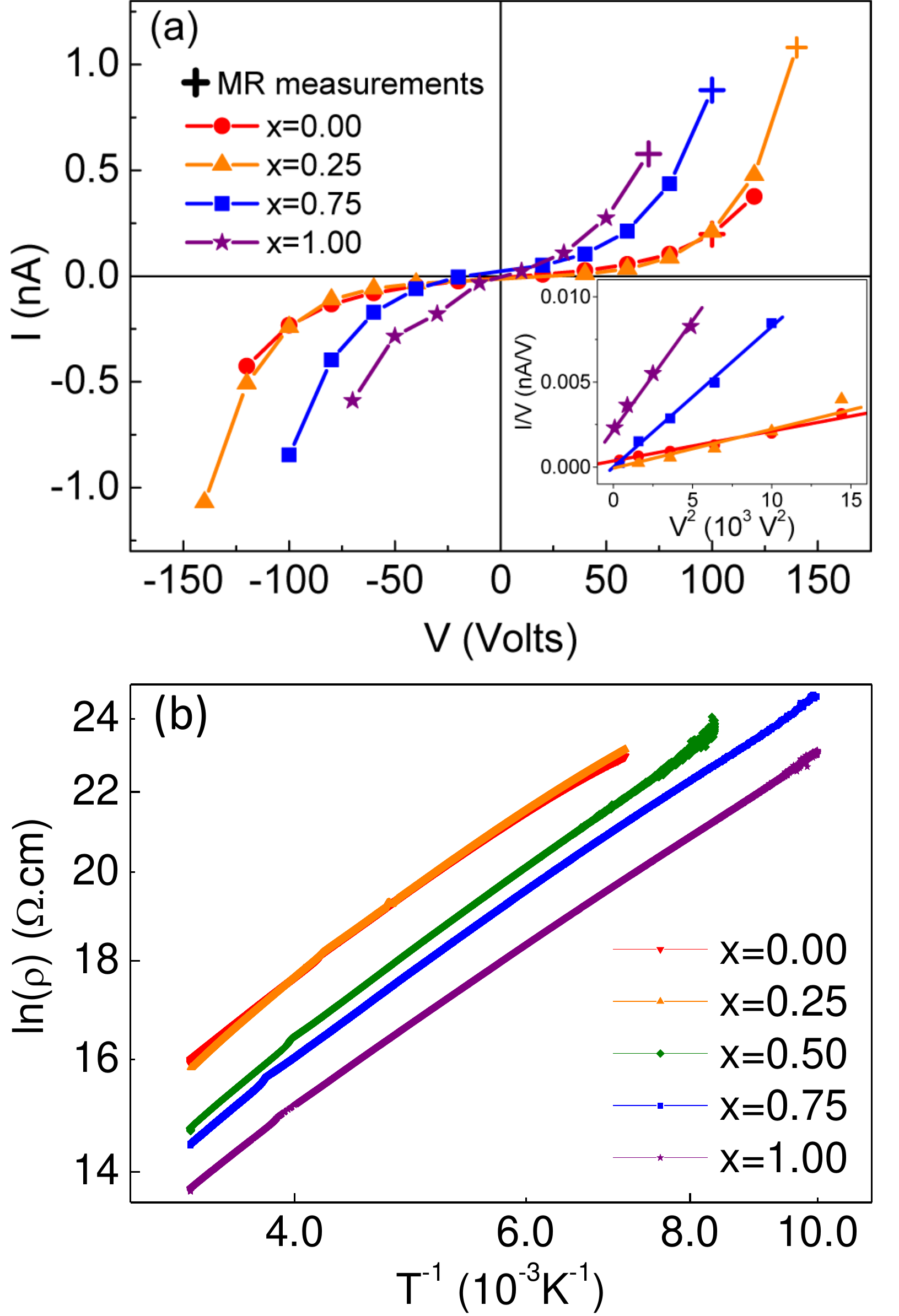}
	\caption{(a) Current-voltage characteristics
		measured at 100 K for all the Fe$_3$O$_4$/Co$_{1-x}$Zn$_x$Fe$_2$O$_4$ annealed devices. The inset shows I/V \textit{versus} V$^2$. (b) Logarithm of the electrical resistivity as a function of  $1/T$, both in logarithmic scale.}
	\label{IVRT}
\end{figure}

\begin{figure*}[ht]
	\includegraphics[width=\textwidth]{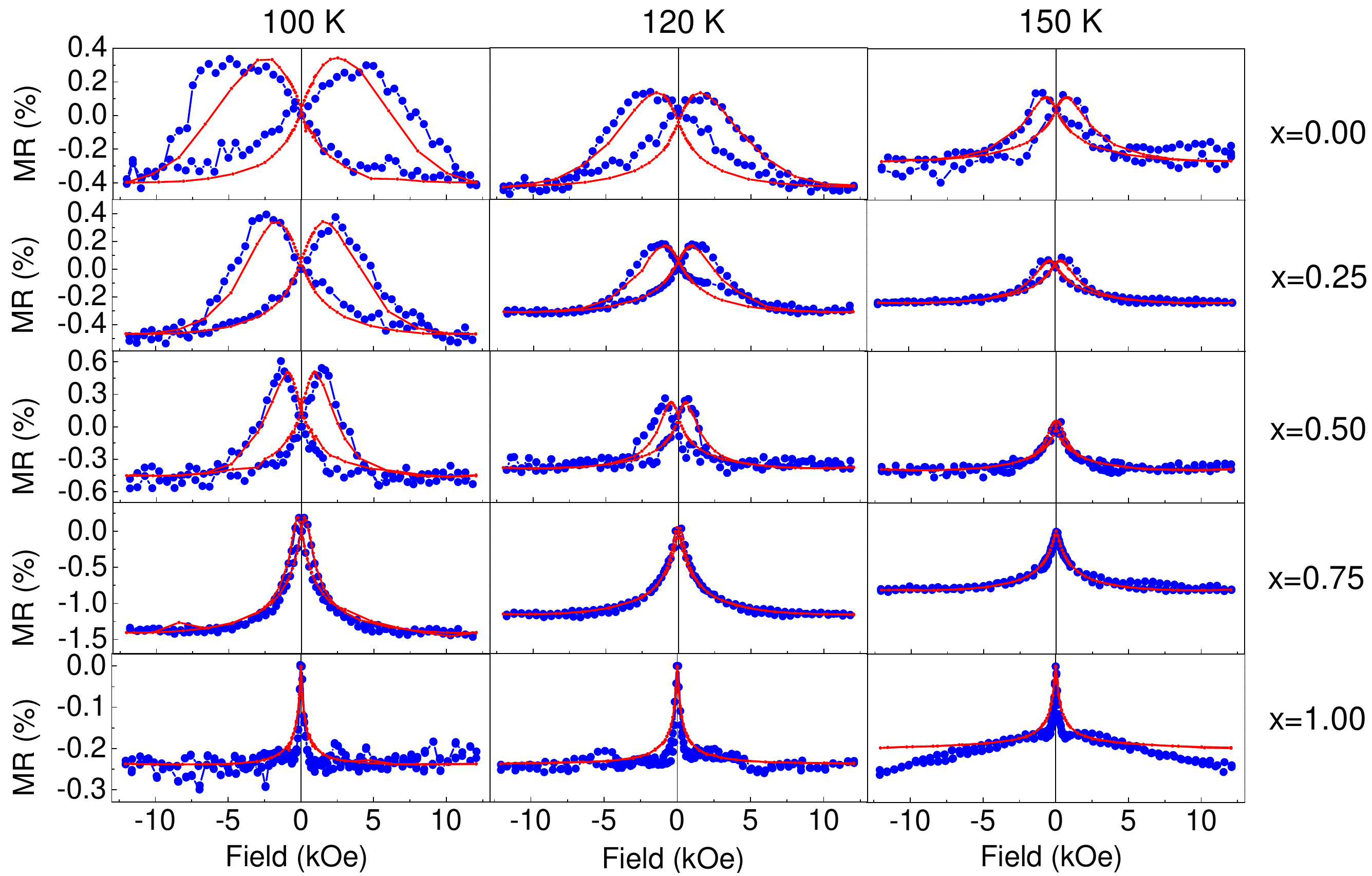}
	\caption{TMR of the devices studied in this work at different
		temperatures. The magnetoresistance is defined as
		$MR=100\times[R(H)-R(0)]/R(0)$, and it was measured with $H$ in the
		plane of the self-assembled films. The red line represents the
		fitting to equation \ref{equationMR} using the experimental $M(H)$ curves shown in Figure \ref{HC},
		measured for each self-assembly of magnetic nanoparticles.}
	\label{MR}
\end{figure*}

The electronic transport in the annealed self-assembled devices was studied from the current-voltage ($I-V$) measurements and from the temperature dependence of the resistivity $\rho$, as reported in Figure \ref{IVRT}. From this figure it can be affirmed that the electron conduction in the devices is given by two independent mechanisms: thermally activated hopping, which is revealed from the temperature dependence of $\rho$; and the tunneling conduction manifested by the non-Ohmic behavior in the I-V curve with the characteristic $V^{3}$ dependence at low temperature, as discussed next. From the Simmons's model, which considers inelastic tunneling across an insulating barrier, the $I-V$ curves can be quantitatively linked to the physical parameters of the system, \textit{i.e.} the tunnel barrier height
($h$) and width ($w$), the effective contact area, etc. \cite{Simmons1963}. This model also considers  linear dependence approximation of the barrier potential profile with
$V$ and $w$. For voltage smaller than the potential barrier, the Simmons's model can be approximated with the well known polynomial relationship:\cite{Simmons1963_V3}

\begin{equation}
\frac{I}{V}\approx G_0\left( 1+ F V^2\right)
\end{equation}
where $G_0$ is  the equilibrium conductance and $F\propto\xi^{2}$, where $\xi=w/h$ is the shape factor of the barrier.\cite{Vilan2007} From the
plot shown in the inset of Figure \ref{IVRT}(a), the shape factor of
the barrier is clearly increasing with $x$. Given that the barrier width is approximately constant for all samples, this reflects a
progressive decrease of the tunneling barrier height as the Zn content increases.
Considering the geometry of the assemblies devices, \textit{i.e.} the area measured by SEM and the film thickness obtained by AFM microscopy, we have plotted $\rho(T)$ in Figure \ref{IVRT}(b), which suggests that a thermally activated transport mechanism is also involved in the conduction of the devices. The temperature dependence follows the relation $ln (\rho) \propto T^{-1/ \alpha}$, where $\alpha=0.4(1)$ was found as the best fitted parameter for all the systems. This value is close to the dependence found in the Efros's variable range hopping model $ln (\rho) \propto T^{-1/2}$, \cite{Efros1975} and it is consistent with the behavior measured in other nanoparticles arrays. \cite{Black2000,Dong2010,Song2013, Kant2008,Zhang2010}.
Moreover, $\rho(T)$ in the core/shell assemblies is, at least, two orders of magnitude smaller than the reported for pellets of  Co$_{1-x}$Zn$_{x}$Fe$_{2}$O$_{4}$ powder ferrite (for example at room temperature $\rho($CoFe$_{2}$O$_{4})\sim$1.10$^7$$\Omega.cm$ and $\rho($Co$_{0.4}$Zn$_{0.6}$Fe$_{2}$O$_{4})\sim$1.10$^8$$\Omega.cm$ \cite{Rani2013,Gul2007,Ramana1999}). This result shows that the tunnel conduction strongly increases the conductivity of the core/shell assemblies compared to the values measured in the semiconductor shell material.  Notice that the larger the tunneling current, the lower the resistivity of the devices.  Therefore, the material that has the lower tunneling barrier will be the one that is most influenced by this contribution.  Consistently, from Figure \ref{IVRT} we have measured that the assemblies of nanoparticles with a ZnFe$_2$O$_4$ shell present the lowest energy barrier, and a systematic increase with decreasing $x$ is observed.

\begin{figure}
    	\includegraphics[width=0.5\textwidth]{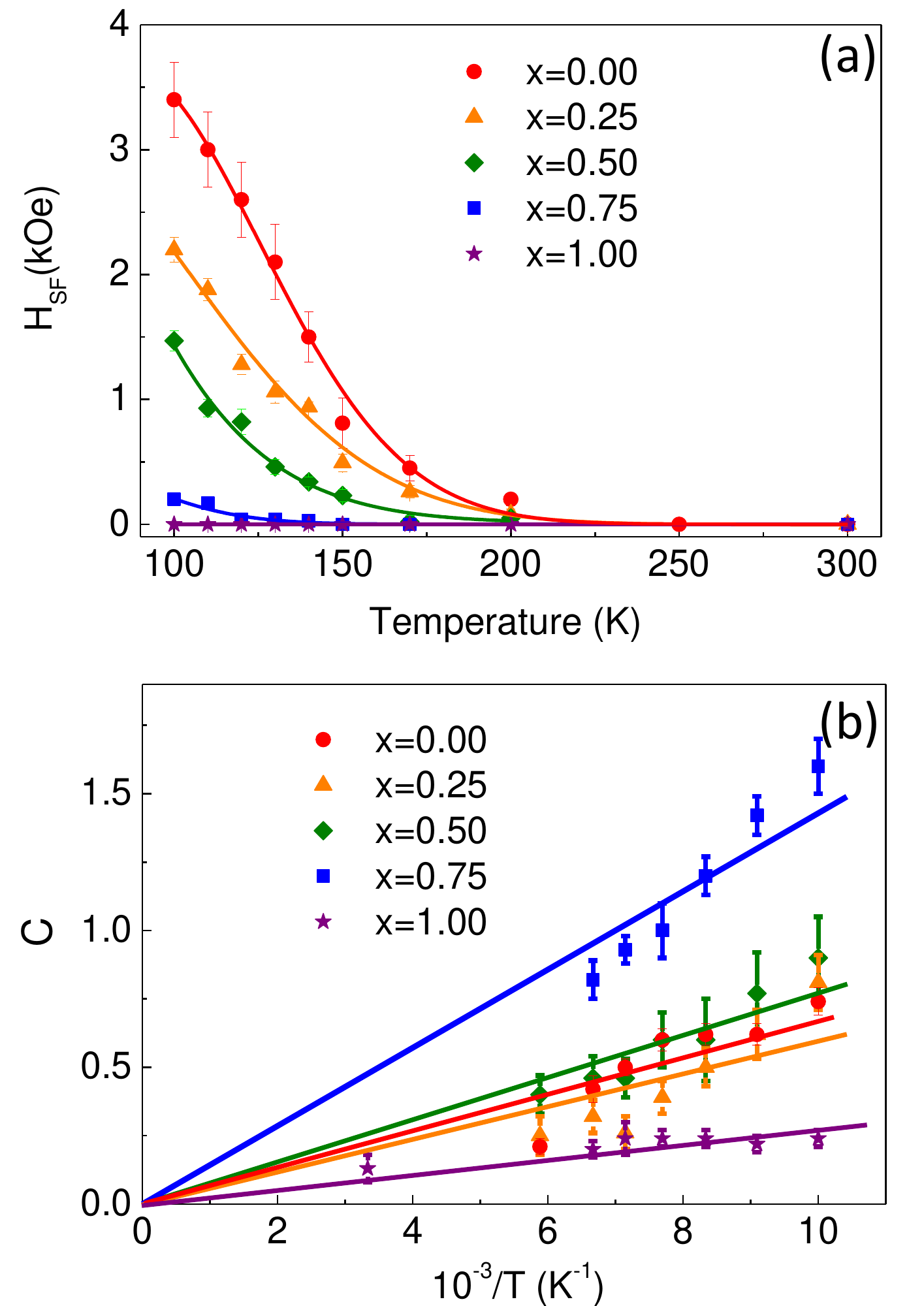}
    
    	\caption{ (a) Switching field ($H_{SF}$) as a function of the temperature for Fe$_3$O$_4$/Co$_{1-x}$Zn$_x$Fe$_2$O$_4$	nanoparticles and (b) $C=JP/4k_BT$ obtained as the adjusted parameter from equation \ref{equationMR} as a function of the inverse of the temperature.}
    	\label{CT}
    \end{figure}

The main findings of this work are summarized in  Figure \ref{MR}  and \ref{CT}. Magnetoresistance curves measured at different temperatures for the five devices studied in this work are shown in  Figure \ref{MR} and Figure S7 of Supplemental Material \cite{SupMat}. At a given temperature, the switching field of the magnetoresistance curve monotonously decreases when the Zn concentration of the shell increases. Moreover, as shown in Figure \ref{CT}, each sample shows a smooth decrease of the switching field with temperature, consistently with the temperature
evolution of the coercive field. Also, a saturation behavior of TMR
at high magnetic field is observed when $x$ decreases, in good
correspondence with the M(H) curves, see Figure \ref{HC}.

For  spin-polarized intergrain tunneling the TMR is related to the macroscopic magnetization.\cite{Lopez2003,Hwang1996,Helman1976} For a ferromagnetic-insulator granular system, the electron tunneling across the insulating barrier was calculated including an additional exchange energy arising from the interaction between the tunneling electron spin and the non-parallel magnetic moment of the neighboring grains. \cite{Helman1976,Barzilai1981}. Assuming that the exchange energy can be expressed in terms of the spin correlation function of two ferromagnetic neighboring grain, the magnetoresistance can be expressed as:
\begin{equation}
TMR=-\frac{JP}{4k_BT} [ m^2(H,T)-m^2(0,T)]
 \label{equationMR}
\end{equation}
where $J$ accounts for the magnetic correlations when the electron tunnels through the insulating barrier and $m=M/M_S$.  Notice that within this model the magnetoresistance does not depend on the total resistivity of the sample. The fittings to this equation  using $C=JP/4k_B T$ as the single adjusted parameter for each temperature, are shown in  Figure \ref{MR}.  The good agreement between both measurements from equation \ref{equationMR} confirms the spin polarized tunnel transport in the present devices, where the $C$ parameter gives the proportionality between two independent experiments, $M(H)$ and $\rho(H)$. Moreover, in agreement with equation \ref{equationMR}, the fittings show that $C$ varies approximately linearly with $1/T$ over the  measured temperature range  (see Figure \ref{CT}(b)). However, although equation \ref{equationMR} adjusts the field and temperature dependence of TMR with the magnetization, due to the complex nature of the present system, it is hard to determine the dependence between the C parameter and the shell stoichiometry. Even though  all the systems have the same Fe$_3$O$_4$ core, the Co$_{1-x}$Zn$_x$Fe$_2$O$_4$ shell was synthesized in a second step which lead to a shell thickness dispersion along the Zn composition, moreover from the synthesis and later thermal treatment the nanostructures could also present dispersion in the nanoparticle coating. These experimental factors could affect the magnetic correlations when the electron tunnels through the insulating magnetic barrier and also the surface Fe$_3$O$_4$ spin polarization. Nevertheless, although it is hard to estimate the evolution of the magnetic correlation with the shell stoichiometry, the magnitude of the interaction can be estimated. Assuming the spin polarization of magnetite, ($P\approx 39 \%$) \cite{Hu2002} a coupling  constant of $J\approx 1.0(0.4)$ meV ($\approx 10$ K) is obtained, comparable to the calculations for the intergrain tunneling transport of manganese perovskites \cite{Balcells1998}. 

The TMR amplitude, which is in the $0.3-1.5 \%$ range
depending on the composition, is similar to other reported values
for self assembled nanoparticles, \cite{Dugay2011, Dong2010,Zhang2010} however, it is
much smaller than the calculated from the Julliere's model:
$TMR=\frac{2P^2}{(1-P^2)}$ on the basis of the spin polarization
values of magnetite. This
reduction may be due to the fact that the tunneling probability decreases 
exponentially with the barrier width. According to Refs.
\cite{Moodera1997, Joo2014} for TMR multilayers, the optimal barrier
thickness is in the $1-1.5$ nm range; however, the tunnel current
between Fe$_3$O$_4$ cores in the self-assembled structure must pass
through the insulator barrier of $\sim$ 3 nm width, which is
composed by the cobalt ferrite shell and the amorphous carbon nanoparticles coating.  On
the other hand, although it is known that the TMR diminishes with
increasing bias voltage,\cite{Coey2010} the high resistance of the magnetic nanoparticles devices
determines the experimental parameters, and high voltage bias  $\sim$
100 V had to be applied to perform the transport
measurements at low temperature. These factors make evident the importance of
optimizing the different stages of the fabrication process in order
to increase the conductivity to produce large amplitude and low switching  field TMR devices based on core/shell magnetic nanoparticles.  However, irrespective of the absolute value of TMR, this study demonstrates that the switching field of TMR can be tailored at will in self-assemblies of exchange coupled core-shell nanoparticles, synthesized by an affordable chemical route.

\section{Conclusion}

In summary, we have fabricated self-assemblies of core/shell nanoparticles with controlled TMR. Particularly, we have shown that the magnetic properties can be finely tuned by changing the shell composition which provides a tool to adjust the TMR switching field.  We shows  whereas that the Fe$_{3}$O$_{4}$ core provides the spin polarized transport, the shell acts as tunnel barrier in the self assembly and also modulates the switching field by tuning the magnetic anisotropy through the interface exchange coupling. This approach shows the feasibility to use assemblies of exchange coupled magnetic nanoparticles in TMR devices, where different combinations of soft/hard and hard/soft core/shell configurations can be envisaged. In this way, combinations of materials can be carefully designed to move across the rigid coupling to exchange bias regime, to design devices with tailored magnetotransport response, which gives a promising base for the design of core/shell nanoparticles based devices for fundamental studies or for spintronic applications.

\begin{acknowledgments}
	The authors thank the staff of the INVAP SE Chemistry Laboratory for the ICP-OES spectrometry measurements.
	The authors also acknowledges financial support of Argentinian governmental agency
	ANPCyT (Project No.PICT-2016-0288 and PICT-2015-0883) and UNCuyo (Project No. 06/C527 and 06/C528). The authors gratefully acknowledge the EU-commission financial support
	under the: H2020-MSCA-RISE-2016 SPICOLOST PROJECT No 734187. F. R. acknowledges financial support of the Ministerio de Econom\'ia y Competitividad of Spain (Project No. MAT2016-80762-R), and Xunta de Galicia (Centro Singular de Investigaci\'on de Galicia accreditation 2016-2019) and the European Union (European Regional Development Fund $-$ ERDF).
\end{acknowledgments}


\begin{thebibliography}{0}%
\makeatletter
\providecommand \@ifxundefined [1]{%
 \@ifx{#1\undefined}
}%
\providecommand \@ifnum [1]{%
 \ifnum #1\expandafter \@firstoftwo
 \else \expandafter \@secondoftwo
 \fi
}%
\providecommand \@ifx [1]{%
 \ifx #1\expandafter \@firstoftwo
 \else \expandafter \@secondoftwo
 \fi
}%
\providecommand \natexlab [1]{#1}%
\providecommand \enquote  [1]{``#1''}%
\providecommand \bibnamefont  [1]{#1}%
\providecommand \bibfnamefont [1]{#1}%
\providecommand \citenamefont [1]{#1}%
\providecommand \href@noop [0]{\@secondoftwo}%
\providecommand \href [0]{\begingroup \@sanitize@url \@href}%
\providecommand \@href[1]{\@@startlink{#1}\@@href}%
\providecommand \@@href[1]{\endgroup#1\@@endlink}%
\providecommand \@sanitize@url [0]{\catcode `\\12\catcode `\$12\catcode
  `\&12\catcode `\#12\catcode `\^12\catcode `\_12\catcode `\%12\relax}%
\providecommand \@@startlink[1]{}%
\providecommand \@@endlink[0]{}%
\providecommand \url  [0]{\begingroup\@sanitize@url \@url }%
\providecommand \@url [1]{\endgroup\@href {#1}{\urlprefix }}%
\providecommand \urlprefix  [0]{URL }%
\providecommand \Eprint [0]{\href }%
\providecommand \doibase [0]{http://dx.doi.org/}%
\providecommand \selectlanguage [0]{\@gobble}%
\providecommand \bibinfo  [0]{\@secondoftwo}%
\providecommand \bibfield  [0]{\@secondoftwo}%
\providecommand \translation [1]{[#1]}%
\providecommand \BibitemOpen [0]{}%
\providecommand \bibitemStop [0]{}%
\providecommand \bibitemNoStop [0]{.\EOS\space}%
\providecommand \EOS [0]{\spacefactor3000\relax}%
\providecommand \BibitemShut  [1]{\csname bibitem#1\endcsname}%
\let\auto@bib@innerbib\@empty
\end{thebibliography}%


\begin{thebibliography}{68}%
		\makeatletter
		\providecommand \@ifxundefined [1]{%
			\@ifx{#1\undefined}
		}%
		\providecommand \@ifnum [1]{%
			\ifnum #1\expandafter \@firstoftwo
			\else \expandafter \@secondoftwo
			\fi
		}%
		\providecommand \@ifx [1]{%
			\ifx #1\expandafter \@firstoftwo
			\else \expandafter \@secondoftwo
			\fi
		}%
		\providecommand \natexlab [1]{#1}%
		\providecommand \enquote  [1]{``#1''}%
		\providecommand \bibnamefont  [1]{#1}%
		\providecommand \bibfnamefont [1]{#1}%
		\providecommand \citenamefont [1]{#1}%
		\providecommand \href@noop [0]{\@secondoftwo}%
		\providecommand \href [0]{\begingroup \@sanitize@url \@href}%
		\providecommand \@href[1]{\@@startlink{#1}\@@href}%
		\providecommand \@@href[1]{\endgroup#1\@@endlink}%
		\providecommand \@sanitize@url [0]{\catcode `\\12\catcode `\$12\catcode
			`\&12\catcode `\#12\catcode `\^12\catcode `\_12\catcode `\%12\relax}%
		\providecommand \@@startlink[1]{}%
		\providecommand \@@endlink[0]{}%
		\providecommand \url  [0]{\begingroup\@sanitize@url \@url }%
		\providecommand \@url [1]{\endgroup\@href {#1}{\urlprefix }}%
		\providecommand \urlprefix  [0]{URL }%
		\providecommand \Eprint [0]{\href }%
		\providecommand \doibase [0]{http://dx.doi.org/}%
		\providecommand \selectlanguage [0]{\@gobble}%
		\providecommand \bibinfo  [0]{\@secondoftwo}%
		\providecommand \bibfield  [0]{\@secondoftwo}%
		\providecommand \translation [1]{[#1]}%
		\providecommand \BibitemOpen [0]{}%
		\providecommand \bibitemStop [0]{}%
		\providecommand \bibitemNoStop [0]{.\EOS\space}%
		\providecommand \EOS [0]{\spacefactor3000\relax}%
		\providecommand \BibitemShut  [1]{\csname bibitem#1\endcsname}%
		\let\auto@bib@innerbib\@empty
		\bibitem [{\citenamefont {Baibich}\ \emph {et~al.}(1988)\citenamefont
			{Baibich}, \citenamefont {Broto}, \citenamefont {Fert}, \citenamefont
			{Van~Dau}, \citenamefont {Petroff}, \citenamefont {Etienne}, \citenamefont
			{Creuzet}, \citenamefont {Friederich},\ and\ \citenamefont
			{Chazelas}}]{Baibich}%
		\BibitemOpen
		\bibfield  {author} {\bibinfo {author} {\bibfnamefont {M.~N.}\ \bibnamefont
				{Baibich}}, \bibinfo {author} {\bibfnamefont {J.~M.}\ \bibnamefont {Broto}},
			\bibinfo {author} {\bibfnamefont {A.}~\bibnamefont {Fert}}, \bibinfo {author}
			{\bibfnamefont {F.~Nguyen}\ \bibnamefont {Van~Dau}}, \bibinfo {author}
			{\bibfnamefont {F.}~\bibnamefont {Petroff}}, \bibinfo {author} {\bibfnamefont
				{P.}~\bibnamefont {Etienne}}, \bibinfo {author} {\bibfnamefont
				{G.}~\bibnamefont {Creuzet}}, \bibinfo {author} {\bibfnamefont
				{A.}~\bibnamefont {Friederich}}, \ and\ \bibinfo {author} {\bibfnamefont
				{J.}~\bibnamefont {Chazelas}},\ }\bibfield  {title} {\enquote {\bibinfo
				{title} {Giant magnetoresistance of (001){F}e/(001){C}r magnetic
					superlattices},}\ }\href {\doibase 10.1103/PhysRevLett.61.2472} {\bibfield
			{journal} {\bibinfo  {journal} {Phys. Rev. Lett.}\ }\textbf {\bibinfo
				{volume} {61}},\ \bibinfo {pages} {2472--2475} (\bibinfo {year}
			{1988})}\BibitemShut {NoStop}%
		\bibitem [{\citenamefont {Binasch}\ \emph {et~al.}(1989)\citenamefont
			{Binasch}, \citenamefont {Gr\"unberg}, \citenamefont {Saurenbach},\ and\
			\citenamefont {Zinn}}]{Binasch}%
		\BibitemOpen
		\bibfield  {author} {\bibinfo {author} {\bibfnamefont {G.}~\bibnamefont
				{Binasch}}, \bibinfo {author} {\bibfnamefont {P.}~\bibnamefont {Gr\"unberg}},
			\bibinfo {author} {\bibfnamefont {F.}~\bibnamefont {Saurenbach}}, \ and\
			\bibinfo {author} {\bibfnamefont {W.}~\bibnamefont {Zinn}},\ }\bibfield
		{title} {\enquote {\bibinfo {title} {Enhanced magnetoresistance in layered
					magnetic structures with antiferromagnetic interlayer exchange},}\ }\href
		{\doibase 10.1103/PhysRevB.39.4828} {\bibfield  {journal} {\bibinfo
				{journal} {Phys. Rev. B}\ }\textbf {\bibinfo {volume} {39}},\ \bibinfo
			{pages} {4828--4830} (\bibinfo {year} {1989})}\BibitemShut {NoStop}%
		\bibitem [{\citenamefont {Zutic}\ \emph {et~al.}(2004)\citenamefont {Zutic},
			\citenamefont {Jaroslav},\ and\ \citenamefont {Sarma}}]{RevModPhys.76.323}%
		\BibitemOpen
		\bibfield  {author} {\bibinfo {author} {\bibfnamefont {I.}~\bibnamefont
				{Zutic}}, \bibinfo {author} {\bibfnamefont {F.}~\bibnamefont {Jaroslav}}, \
			and\ \bibinfo {author} {\bibfnamefont {S.~Das}\ \bibnamefont {Sarma}},\
		}\bibfield  {title} {\enquote {\bibinfo {title} {Spintronics: Fundamentals
					and applications},}\ }\href {\doibase 10.1103/RevModPhys.76.323} {\bibfield
			{journal} {\bibinfo  {journal} {Rev. Mod. Phys.}\ }\textbf {\bibinfo {volume}
				{76}},\ \bibinfo {pages} {323--410} (\bibinfo {year} {2004})}\BibitemShut
		{NoStop}%
		\bibitem [{\citenamefont {Bader}\ and\ \citenamefont
			{Parkin}(2010)}]{Parkin2010}%
		\BibitemOpen
		\bibfield  {author} {\bibinfo {author} {\bibfnamefont {S.D.}\ \bibnamefont
				{Bader}}\ and\ \bibinfo {author} {\bibfnamefont {S.S.P.}\ \bibnamefont
				{Parkin}},\ }\bibfield  {title} {\enquote {\bibinfo {title} {Spintronics},}\
		}\href {\doibase 10.1146/annurev-conmatphys-070909-104123} {\bibfield
			{journal} {\bibinfo  {journal} {Annual Review of Condensed Matter Physics}\
			}\textbf {\bibinfo {volume} {1}},\ \bibinfo {pages} {71--88} (\bibinfo {year}
			{2010})}\BibitemShut {NoStop}%
		\bibitem [{\citenamefont {Julliere}(1975)}]{JULLIERE1975225}%
		\BibitemOpen
		\bibfield  {author} {\bibinfo {author} {\bibfnamefont {M.}~\bibnamefont
				{Julliere}},\ }\bibfield  {title} {\enquote {\bibinfo {title} {Tunneling
					between ferromagnetic films},}\ }\href {\doibase
			https://doi.org/10.1016/0375-9601(75)90174-7} {\bibfield  {journal} {\bibinfo
				{journal} {Physics Letters A}\ }\textbf {\bibinfo {volume} {54}},\ \bibinfo
			{pages} {225 -- 226} (\bibinfo {year} {1975})}\BibitemShut {NoStop}%
		\bibitem [{\citenamefont {Moodera}\ \emph {et~al.}(1995)\citenamefont
			{Moodera}, \citenamefont {Kinder}, \citenamefont {Wong},\ and\ \citenamefont
			{Meservey}}]{Moodera}%
		\BibitemOpen
		\bibfield  {author} {\bibinfo {author} {\bibfnamefont {J.~S.}\ \bibnamefont
				{Moodera}}, \bibinfo {author} {\bibfnamefont {L.~R.}\ \bibnamefont {Kinder}},
			\bibinfo {author} {\bibfnamefont {T.~M.}\ \bibnamefont {Wong}}, \ and\
			\bibinfo {author} {\bibfnamefont {R.}~\bibnamefont {Meservey}},\ }\bibfield
		{title} {\enquote {\bibinfo {title} {Large magnetoresistance at room
					temperature in ferromagnetic thin film tunnel junctions},}\ }\href {\doibase
			10.1103/PhysRevLett.74.3273} {\bibfield  {journal} {\bibinfo  {journal}
				{Phys. Rev. Lett.}\ }\textbf {\bibinfo {volume} {74}},\ \bibinfo {pages}
			{3273--3276} (\bibinfo {year} {1995})}\BibitemShut {NoStop}%
		\bibitem [{\citenamefont {Yuasa}\ \emph {et~al.}(2004)\citenamefont {Yuasa},
			\citenamefont {Nagahama}, \citenamefont {Fukushima}, \citenamefont {Suzuki},\
			and\ \citenamefont {Ando}}]{Yuasa2004}%
		\BibitemOpen
		\bibfield  {author} {\bibinfo {author} {\bibfnamefont {S.}~\bibnamefont
				{Yuasa}}, \bibinfo {author} {\bibfnamefont {T.}~\bibnamefont {Nagahama}},
			\bibinfo {author} {\bibfnamefont {A.}~\bibnamefont {Fukushima}}, \bibinfo
			{author} {\bibfnamefont {Y.}~\bibnamefont {Suzuki}}, \ and\ \bibinfo {author}
			{\bibfnamefont {K.}~\bibnamefont {Ando}},\ }\bibfield  {title} {\enquote
			{\bibinfo {title} {Giant room-temperature magnetoresistance in single-crystal
					{Fe/MgO/Fe} magnetic tunnel junctions},}\ }\href {\doibase 10.1038/nmat1257}
		{\bibfield  {journal} {\bibinfo  {journal} {Nature Materials}\ }\textbf
			{\bibinfo {volume} {3}},\ \bibinfo {pages} {868} (\bibinfo {year}
			{2004})}\BibitemShut {NoStop}%
		\bibitem [{\citenamefont {Djayaprawira}\ \emph {et~al.}(2005)\citenamefont
			{Djayaprawira}, \citenamefont {Tsunekawa}, \citenamefont {Nagai},
			\citenamefont {Maehara}, \citenamefont {Yamagata}, \citenamefont {Watanabe},
			\citenamefont {Yuasa}, \citenamefont {Suzuki},\ and\ \citenamefont
			{Ando}}]{Djayaprawira2005}%
		\BibitemOpen
		\bibfield  {author} {\bibinfo {author} {\bibfnamefont {D.~D.}\ \bibnamefont
				{Djayaprawira}}, \bibinfo {author} {\bibfnamefont {K.}~\bibnamefont
				{Tsunekawa}}, \bibinfo {author} {\bibfnamefont {M.}~\bibnamefont {Nagai}},
			\bibinfo {author} {\bibfnamefont {H.}~\bibnamefont {Maehara}}, \bibinfo
			{author} {\bibfnamefont {S.}~\bibnamefont {Yamagata}}, \bibinfo {author}
			{\bibfnamefont {N.}~\bibnamefont {Watanabe}}, \bibinfo {author}
			{\bibfnamefont {S.}~\bibnamefont {Yuasa}}, \bibinfo {author} {\bibfnamefont
				{Y.}~\bibnamefont {Suzuki}}, \ and\ \bibinfo {author} {\bibfnamefont
				{K.}~\bibnamefont {Ando}},\ }\bibfield  {title} {\enquote {\bibinfo {title}
				{230\% room-temperature magnetoresistance in {CoFeB/MgO/CoFeB} magnetic
					tunnel junctions},}\ }\href {\doibase 10.1063/1.1871344} {\bibfield
			{journal} {\bibinfo  {journal} {Applied Physics Letters}\ }\textbf {\bibinfo
				{volume} {86}},\ \bibinfo {pages} {092502} (\bibinfo {year}
			{2005})}\BibitemShut {NoStop}%
		\bibitem [{\citenamefont {L\'{o}pez-Quintela}\ \emph
			{et~al.}(2003)\citenamefont {L\'{o}pez-Quintela}, \citenamefont {Hueso},
			\citenamefont {Rivas},\ and\ \citenamefont {Rivadulla}}]{Lopez2003}%
		\BibitemOpen
		\bibfield  {author} {\bibinfo {author} {\bibfnamefont {M.~A.}\ \bibnamefont
				{L\'{o}pez-Quintela}}, \bibinfo {author} {\bibfnamefont {L.~E.}\ \bibnamefont
				{Hueso}}, \bibinfo {author} {\bibfnamefont {J.}~\bibnamefont {Rivas}}, \ and\
			\bibinfo {author} {\bibfnamefont {F.}~\bibnamefont {Rivadulla}},\ }\bibfield
		{title} {\enquote {\bibinfo {title} {Intergranular magnetoresistance in
					nanomanganites},}\ }\href {http://stacks.iop.org/0957-4484/14/i=2/a=322}
		{\bibfield  {journal} {\bibinfo  {journal} {Nanotechnology}\ }\textbf
			{\bibinfo {volume} {14}},\ \bibinfo {pages} {212} (\bibinfo {year}
			{2003})}\BibitemShut {NoStop}%
		\bibitem [{\citenamefont {Hwang}\ \emph {et~al.}(1996)\citenamefont {Hwang},
			\citenamefont {Cheong}, \citenamefont {Ong},\ and\ \citenamefont
			{Batlogg}}]{Hwang1996}%
		\BibitemOpen
		\bibfield  {author} {\bibinfo {author} {\bibfnamefont {H.~Y.}\ \bibnamefont
				{Hwang}}, \bibinfo {author} {\bibfnamefont {S-W.}\ \bibnamefont {Cheong}},
			\bibinfo {author} {\bibfnamefont {N.~P.}\ \bibnamefont {Ong}}, \ and\
			\bibinfo {author} {\bibfnamefont {B.}~\bibnamefont {Batlogg}},\ }\bibfield
		{title} {\enquote {\bibinfo {title} {Spin-polarized intergrain tunneling in
					{La$_{2/3}$Sr$_{1/3}$MnO$_3$}},}\ }\href
		{https://journals.aps.org/prl/pdf/10.1103/PhysRevLett.77.2041} {\bibfield
			{journal} {\bibinfo  {journal} {Physical Review Letters}\ }\textbf {\bibinfo
				{volume} {77}},\ \bibinfo {pages} {2041} (\bibinfo {year}
			{1996})}\BibitemShut {NoStop}%
		\bibitem [{\citenamefont {Murray}\ \emph {et~al.}(2000)\citenamefont {Murray},
			\citenamefont {Kagan},\ and\ \citenamefont {Bawendi}}]{Murray2000}%
		\BibitemOpen
		\bibfield  {author} {\bibinfo {author} {\bibfnamefont {C.B.}\ \bibnamefont
				{Murray}}, \bibinfo {author} {\bibfnamefont {C.R.}\ \bibnamefont {Kagan}}, \
			and\ \bibinfo {author} {\bibfnamefont {M.G.}\ \bibnamefont {Bawendi}},\
		}\bibfield  {title} {\enquote {\bibinfo {title} {Synthesis and
					characterization of monodisperse nanocrystals and close-packed nanocrystal
					assemblies},}\ }\href {\doibase 10.1146/annurev.matsci.30.1.545} {\bibfield
			{journal} {\bibinfo  {journal} {Annual Review of Materials Science}\ }\textbf
			{\bibinfo {volume} {30}},\ \bibinfo {pages} {545--610} (\bibinfo {year}
			{2000})}\BibitemShut {NoStop}%
		\bibitem [{\citenamefont {Sun}\ \emph {et~al.}(2000)\citenamefont {Sun},
			\citenamefont {Murray}, \citenamefont {Weller}, \citenamefont {Folks},\ and\
			\citenamefont {Moser}}]{Sun2000}%
		\BibitemOpen
		\bibfield  {author} {\bibinfo {author} {\bibfnamefont {S.}~\bibnamefont
				{Sun}}, \bibinfo {author} {\bibfnamefont {C.~B.}\ \bibnamefont {Murray}},
			\bibinfo {author} {\bibfnamefont {D.}~\bibnamefont {Weller}}, \bibinfo
			{author} {\bibfnamefont {L.}~\bibnamefont {Folks}}, \ and\ \bibinfo {author}
			{\bibfnamefont {A.}~\bibnamefont {Moser}},\ }\bibfield  {title} {\enquote
			{\bibinfo {title} {{Monodisperse FePt Nanoparticles and Ferromagnetic FePt
						Nanocrystal Superlattices}},}\ }\href {\doibase
			10.1126/science.287.5460.1989} {\bibfield  {journal} {\bibinfo  {journal}
				{{Science}}\ }\textbf {\bibinfo {volume} {287}},\ \bibinfo {pages} {1989--92}
			(\bibinfo {year} {2000})}\BibitemShut {NoStop}%
		\bibitem [{\citenamefont {Zeng}\ \emph {et~al.}(2006)\citenamefont {Zeng},
			\citenamefont {Black}, \citenamefont {Sandstrom}, \citenamefont {Rice},
			\citenamefont {Murray},\ and\ \citenamefont {Sun}}]{Zeng2006}%
		\BibitemOpen
		\bibfield  {author} {\bibinfo {author} {\bibfnamefont {H.}~\bibnamefont
				{Zeng}}, \bibinfo {author} {\bibfnamefont {C.~T.}\ \bibnamefont {Black}},
			\bibinfo {author} {\bibfnamefont {R.~L.}\ \bibnamefont {Sandstrom}}, \bibinfo
			{author} {\bibfnamefont {P.~M.}\ \bibnamefont {Rice}}, \bibinfo {author}
			{\bibfnamefont {C.~B.}\ \bibnamefont {Murray}}, \ and\ \bibinfo {author}
			{\bibfnamefont {Shouheng}\ \bibnamefont {Sun}},\ }\bibfield  {title}
		{\enquote {\bibinfo {title} {Magnetotransport of magnetite nanoparticle
					arrays},}\ }\href {\doibase 10.1103/PhysRevB.73.020402} {\bibfield  {journal}
			{\bibinfo  {journal} {Phys. Rev. B}\ }\textbf {\bibinfo {volume} {73}},\
			\bibinfo {pages} {020402} (\bibinfo {year} {2006})}\BibitemShut {NoStop}%
		\bibitem [{\citenamefont {Black}\ \emph {et~al.}(2000)\citenamefont {Black},
			\citenamefont {Murray}, \citenamefont {Sandstrom},\ and\ \citenamefont
			{Sun}}]{Black2000}%
		\BibitemOpen
		\bibfield  {author} {\bibinfo {author} {\bibfnamefont {C.~T.}\ \bibnamefont
				{Black}}, \bibinfo {author} {\bibfnamefont {C.~B.}\ \bibnamefont {Murray}},
			\bibinfo {author} {\bibfnamefont {R.~L.}\ \bibnamefont {Sandstrom}}, \ and\
			\bibinfo {author} {\bibfnamefont {S.}~\bibnamefont {Sun}},\ }\bibfield
		{title} {\enquote {\bibinfo {title} {Spin-dependent tunneling in
					self-assembled cobalt-nanocrystal superlattices},}\ }\href {\doibase
			10.1126/science.290.5494.1131} {\ \textbf {\bibinfo {volume} {290}},\
			\bibinfo {pages} {1131--1134} (\bibinfo {year} {2000})}\BibitemShut {NoStop}%
		\bibitem [{\citenamefont {Wang}\ \emph {et~al.}(2009)\citenamefont {Wang},
			\citenamefont {Yue}, \citenamefont {Wua}, \citenamefont {Zhang},
			\citenamefont {Zhong},\ and\ \citenamefont {Du}}]{Wang2009}%
		\BibitemOpen
		\bibfield  {author} {\bibinfo {author} {\bibfnamefont {S.}~\bibnamefont
				{Wang}}, \bibinfo {author} {\bibfnamefont {F.~J.}\ \bibnamefont {Yue}},
			\bibinfo {author} {\bibfnamefont {D.}~\bibnamefont {Wua}}, \bibinfo {author}
			{\bibfnamefont {F.~M.}\ \bibnamefont {Zhang}}, \bibinfo {author}
			{\bibfnamefont {W.}~\bibnamefont {Zhong}}, \ and\ \bibinfo {author}
			{\bibfnamefont {Y.~W.}\ \bibnamefont {Du}},\ }\bibfield  {title} {\enquote
			{\bibinfo {title} {Enhanced magnetoresistance in self-assembled monolayer of
					oleic acid molecules on {Fe$_3$O$_4$} nanoparticles},}\ }\href
		{https://doi.org/10.1063/1.3059571} {\bibfield  {journal} {\bibinfo
				{journal} {Applied Physics Letters}\ }\textbf {\bibinfo {volume} {94}},\
			\bibinfo {pages} {012507} (\bibinfo {year} {2009})}\BibitemShut {NoStop}%
		\bibitem [{\citenamefont {Dugay}\ \emph {et~al.}(2011)\citenamefont {Dugay},
			\citenamefont {Tan}, \citenamefont {Meffre}, \citenamefont {Blon},
			\citenamefont {Lacroix}, \citenamefont {Carrey}, \citenamefont {Fazzini},
			\citenamefont {Lachaize}, \citenamefont {Chaudret},\ and\ \citenamefont
			{Respaud}}]{Dugay2011}%
		\BibitemOpen
		\bibfield  {author} {\bibinfo {author} {\bibfnamefont {J.}~\bibnamefont
				{Dugay}}, \bibinfo {author} {\bibfnamefont {R.~P.}\ \bibnamefont {Tan}},
			\bibinfo {author} {\bibfnamefont {A.}~\bibnamefont {Meffre}}, \bibinfo
			{author} {\bibfnamefont {T.}~\bibnamefont {Blon}}, \bibinfo {author}
			{\bibfnamefont {L.~M.}\ \bibnamefont {Lacroix}}, \bibinfo {author}
			{\bibfnamefont {J.}~\bibnamefont {Carrey}}, \bibinfo {author} {\bibfnamefont
				{P.~F.}\ \bibnamefont {Fazzini}}, \bibinfo {author} {\bibfnamefont
				{S.}~\bibnamefont {Lachaize}}, \bibinfo {author} {\bibfnamefont
				{B.}~\bibnamefont {Chaudret}}, \ and\ \bibinfo {author} {\bibfnamefont
				{M.}~\bibnamefont {Respaud}},\ }\bibfield  {title} {\enquote {\bibinfo
				{title} {Room-temperature tunnel magnetoresistance in self-assembled
					chemically synthesized metallic iron nanoparticles},}\ }\href {\doibase
			10.1021/nl203284v} {\bibfield  {journal} {\bibinfo  {journal} {Nano Letters}\
			}\textbf {\bibinfo {volume} {11}},\ \bibinfo {pages} {5128--5134} (\bibinfo
			{year} {2011})}\BibitemShut {NoStop}%
		\bibitem [{\citenamefont {Mohan~Kant}\ \emph {et~al.}(2008)\citenamefont
			{Mohan~Kant}, \citenamefont {Sethupathi},\ and\ \citenamefont
			{Ramachandra~Rao}}]{Kant2008}%
		\BibitemOpen
		\bibfield  {author} {\bibinfo {author} {\bibfnamefont {K.}~\bibnamefont
				{Mohan~Kant}}, \bibinfo {author} {\bibfnamefont {K.}~\bibnamefont
				{Sethupathi}}, \ and\ \bibinfo {author} {\bibfnamefont {M.~S.}\ \bibnamefont
				{Ramachandra~Rao}},\ }\bibfield  {title} {\enquote {\bibinfo {title} {Role of
					oxide barrier in intergranular tunnel junctions: An enhanced
					magnetoresistance in {SiO$_{2}$} and {ZnO} coated {Fe$_{3}$O$_{4}$}
					nanoparticle compacts},}\ }\href {\doibase 10.1063/1.2840902} {\bibfield
			{journal} {\bibinfo  {journal} {Journal of Applied Physics}\ }\textbf
			{\bibinfo {volume} {103}},\ \bibinfo {pages} {07F318} (\bibinfo {year}
			{2008})}\BibitemShut {NoStop}%
		\bibitem [{\citenamefont {Taub}\ \emph {et~al.}(2009)\citenamefont {Taub},
			\citenamefont {Tsukernik},\ and\ \citenamefont {Markovich}}]{Taub2009}%
		\BibitemOpen
		\bibfield  {author} {\bibinfo {author} {\bibfnamefont {N.}~\bibnamefont
				{Taub}}, \bibinfo {author} {\bibfnamefont {A.}~\bibnamefont {Tsukernik}}, \
			and\ \bibinfo {author} {\bibfnamefont {G.}~\bibnamefont {Markovich}},\
		}\bibfield  {title} {\enquote {\bibinfo {title} {Interparticle spin-polarized
					tunneling in arrays of magnetite nanocrystals},}\ }\href {\doibase
			https://doi.org/10.1016/j.jmmm.2008.12.012} {\bibfield  {journal} {\bibinfo
				{journal} {Journal of Magnetism and Magnetic Materials}\ }\textbf {\bibinfo
				{volume} {321}},\ \bibinfo {pages} {1933} (\bibinfo {year}
			{2009})}\BibitemShut {NoStop}%
		\bibitem [{\citenamefont {Tran}\ \emph {et~al.}(2008)\citenamefont {Tran},
			\citenamefont {Beloborodov}, \citenamefont {Hu}, \citenamefont {Lin},
			\citenamefont {Rosenbaum},\ and\ \citenamefont {Jaeger}}]{Tran2008}%
		\BibitemOpen
		\bibfield  {author} {\bibinfo {author} {\bibfnamefont {T.~B.}\ \bibnamefont
				{Tran}}, \bibinfo {author} {\bibfnamefont {I.~S.}\ \bibnamefont
				{Beloborodov}}, \bibinfo {author} {\bibfnamefont {Jingshi}\ \bibnamefont
				{Hu}}, \bibinfo {author} {\bibfnamefont {X.~M.}\ \bibnamefont {Lin}},
			\bibinfo {author} {\bibfnamefont {T.~F.}\ \bibnamefont {Rosenbaum}}, \ and\
			\bibinfo {author} {\bibfnamefont {H.~M.}\ \bibnamefont {Jaeger}},\ }\bibfield
		{title} {\enquote {\bibinfo {title} {Sequential tunneling and inelastic
					cotunneling in nanoparticle arrays},}\ }\href {\doibase
			10.1103/PhysRevB.78.075437} {\bibfield  {journal} {\bibinfo  {journal} {Phys.
					Rev. B}\ }\textbf {\bibinfo {volume} {78}},\ \bibinfo {pages} {075437}
			(\bibinfo {year} {2008})}\BibitemShut {NoStop}%
		\bibitem [{\citenamefont {Chen}\ \emph {et~al.}(2013)\citenamefont {Chen},
			\citenamefont {Ye}, \citenamefont {Soong}, \citenamefont {Kikkawa},
			\citenamefont {Kagan},\ and\ \citenamefont {Murray}}]{Chen2013}%
		\BibitemOpen
		\bibfield  {author} {\bibinfo {author} {\bibfnamefont {J.}~\bibnamefont
				{Chen}}, \bibinfo {author} {\bibfnamefont {X.}~\bibnamefont {Ye}}, \bibinfo
			{author} {\bibfnamefont {J.~O.}\ \bibnamefont {Soong}}, \bibinfo {author}
			{\bibfnamefont {J.~M.}\ \bibnamefont {Kikkawa}}, \bibinfo {author}
			{\bibfnamefont {C.~R.}\ \bibnamefont {Kagan}}, \ and\ \bibinfo {author}
			{\bibfnamefont {C.~B.}\ \bibnamefont {Murray}},\ }\bibfield  {title}
		{\enquote {\bibinfo {title} {{Bistable Magnetoresistance Switching Binary
						Nanocrystal Superlattices by Self-Assembly and Thermal Annealing}},}\ }\href
		{\doibase 10.1021/nn3052617} {\bibfield  {journal} {\bibinfo  {journal} {ACS
					Nano}\ }\textbf {\bibinfo {volume} {7}},\ \bibinfo {pages} {1478--1486}
			(\bibinfo {year} {2013})}\BibitemShut {NoStop}%
		\bibitem [{\citenamefont {Dong}\ \emph {et~al.}(2010)\citenamefont {Dong},
			\citenamefont {Chen}, \citenamefont {Vora},\ and\ \citenamefont
			{Murray}}]{Dong2010}%
		\BibitemOpen
		\bibfield  {author} {\bibinfo {author} {\bibfnamefont {A.}~\bibnamefont
				{Dong}}, \bibinfo {author} {\bibfnamefont {J.}~\bibnamefont {Chen}}, \bibinfo
			{author} {\bibfnamefont {J.~M.}\ \bibnamefont {Vora}, \bibfnamefont
				{P.~M.and~Kikkawa}}, \ and\ \bibinfo {author} {\bibfnamefont {C.B.}\
				\bibnamefont {Murray}},\ }\bibfield  {title} {\enquote {\bibinfo {title}
				{{Binary Nanocrystal Superlattice Membranes Self-Assembled at the Liquid-Air
						Interface}},}\ }\href {\doibase 10.1038/nature09188} {\bibfield  {journal}
			{\bibinfo  {journal} {Nature}\ }\textbf {\bibinfo {volume} {466}},\ \bibinfo
			{pages} {474--477} (\bibinfo {year} {2010})}\BibitemShut {NoStop}%
		\bibitem [{\citenamefont {Jiang}\ \emph {et~al.}(2017)\citenamefont {Jiang},
			\citenamefont {Leung},\ and\ \citenamefont {Pong}}]{Jiang2017}%
		\BibitemOpen
		\bibfield  {author} {\bibinfo {author} {\bibfnamefont {C.}~\bibnamefont
				{Jiang}}, \bibinfo {author} {\bibfnamefont {C.~W.}\ \bibnamefont {Leung}}, \
			and\ \bibinfo {author} {\bibfnamefont {P.~W.T.}\ \bibnamefont {Pong}},\
		}\bibfield  {title} {\enquote {\bibinfo {title} {Self-assembled thin films of
					{Fe$_3$O$_4$-Ag} composite nanoparticles for spintronic applications},}\
		}\href {\doibase https://doi.org/10.1016/j.apsusc.2017.05.116} {\bibfield
			{journal} {\bibinfo  {journal} {Applied Surface Science}\ }\textbf {\bibinfo
				{volume} {419}},\ \bibinfo {pages} {692 -- 696} (\bibinfo {year}
			{2017})}\BibitemShut {NoStop}%
		\bibitem [{\citenamefont {Kumar}\ \emph {et~al.}(2013)\citenamefont {Kumar},
			\citenamefont {Ray}, \citenamefont {Chakraverty},\ and\ \citenamefont
			{Sarma}}]{Kumar2013}%
		\BibitemOpen
		\bibfield  {author} {\bibinfo {author} {\bibfnamefont {P.~A.}\ \bibnamefont
				{Kumar}}, \bibinfo {author} {\bibfnamefont {S.}~\bibnamefont {Ray}}, \bibinfo
			{author} {\bibfnamefont {S.}~\bibnamefont {Chakraverty}}, \ and\ \bibinfo
			{author} {\bibfnamefont {D.~D.}\ \bibnamefont {Sarma}},\ }\bibfield  {title}
		{\enquote {\bibinfo {title} {Engineered spin-valve type magnetoresistance in
					{Fe$_3$O$_4$-CoFe$_2$O$_4$} core$-$shell nanoparticles},}\ }\href {\doibase
			10.1063/1.4819956 View} {\bibfield  {journal} {\bibinfo  {journal} {Appl.
					Phys. Lett.}\ }\textbf {\bibinfo {volume} {103}},\ \bibinfo {pages} {102406}
			(\bibinfo {year} {2013})}\BibitemShut {NoStop}%
		\bibitem [{\citenamefont {Zhang}\ \emph {et~al.}(2010)\citenamefont {Zhang},
			\citenamefont {Xing}, \citenamefont {Poudyal}, \citenamefont {Nandwana},
			\citenamefont {Rong}, \citenamefont {Yan}, \citenamefont {Zeng},\ and\
			\citenamefont {Liu}}]{Zhang2010}%
		\BibitemOpen
		\bibfield  {author} {\bibinfo {author} {\bibfnamefont {Y.}~\bibnamefont
				{Zhang}}, \bibinfo {author} {\bibfnamefont {H.}~\bibnamefont {Xing}},
			\bibinfo {author} {\bibfnamefont {N.}~\bibnamefont {Poudyal}}, \bibinfo
			{author} {\bibfnamefont {V.}~\bibnamefont {Nandwana}}, \bibinfo {author}
			{\bibfnamefont {C.~B.}\ \bibnamefont {Rong}}, \bibinfo {author}
			{\bibfnamefont {S.~S.}\ \bibnamefont {Yan}}, \bibinfo {author} {\bibfnamefont
				{H.}~\bibnamefont {Zeng}}, \ and\ \bibinfo {author} {\bibfnamefont {J.~P.}\
				\bibnamefont {Liu}},\ }\bibfield  {title} {\enquote {\bibinfo {title}
				{Inversed tunneling magnetoresistance in hybrid {FePt/Fe$_{3}$O$_{4}$}
					core/shell nanoparticles systems},}\ }\href {\doibase 10.1063/1.3505778}
		{\bibfield  {journal} {\bibinfo  {journal} {Journal of Applied Physics}\
			}\textbf {\bibinfo {volume} {108}},\ \bibinfo {pages} {103905} (\bibinfo
			{year} {2010})}\BibitemShut {NoStop}%
		\bibitem [{\citenamefont {Helman}\ and\ \citenamefont
			{Abeles}(1976)}]{Helman1976}%
		\BibitemOpen
		\bibfield  {author} {\bibinfo {author} {\bibfnamefont {J.~S.}\ \bibnamefont
				{Helman}}\ and\ \bibinfo {author} {\bibfnamefont {B.}~\bibnamefont
				{Abeles}},\ }\bibfield  {title} {\enquote {\bibinfo {title} {Tunneling of
					spin-polarized electrons and magnetoresistance in granular ni films},}\
		}\href {\doibase 10.1103/PhysRevLett.37.1429} {\bibfield  {journal} {\bibinfo
				{journal} {Phys. Rev. Lett.}\ }\textbf {\bibinfo {volume} {37}},\ \bibinfo
			{pages} {1429--1432} (\bibinfo {year} {1976})}\BibitemShut {NoStop}%
		\bibitem [{\citenamefont {El-Hilo}\ \emph {et~al.}(1998)\citenamefont
			{El-Hilo}, \citenamefont {Chantrell},\ and\ \citenamefont
			{O’Grady}}]{Elhilo1998}%
		\BibitemOpen
		\bibfield  {author} {\bibinfo {author} {\bibfnamefont {M.}~\bibnamefont
				{El-Hilo}}, \bibinfo {author} {\bibfnamefont {R.~W.}\ \bibnamefont
				{Chantrell}}, \ and\ \bibinfo {author} {\bibfnamefont {K.}~\bibnamefont
				{O’Grady}},\ }\bibfield  {title} {\enquote {\bibinfo {title} {A model of
					interaction effects in granular magnetic solids},}\ }\href {\doibase
			10.1063/1.368761} {\bibfield  {journal} {\bibinfo  {journal} {Journal of
					Applied Physics}\ }\textbf {\bibinfo {volume} {84}},\ \bibinfo {pages}
			{5114--5122} (\bibinfo {year} {1998})}\BibitemShut {NoStop}%
		\bibitem [{\citenamefont {Hu}\ and\ \citenamefont {Suzuki}(2002)}]{Hu2002}%
		\BibitemOpen
		\bibfield  {author} {\bibinfo {author} {\bibfnamefont {G.}~\bibnamefont
				{Hu}}\ and\ \bibinfo {author} {\bibfnamefont {Y.}~\bibnamefont {Suzuki}},\
		}\bibfield  {title} {\enquote {\bibinfo {title} {Negative spin polarization
					of fe$_3$o$_4$ in magnetite$/$manganite$-$based junctions},}\ }\href
		{\doibase 10.1103/PhysRevLett.89.276601} {\bibfield  {journal} {\bibinfo
				{journal} {Phys. Rev. Lett.}\ }\textbf {\bibinfo {volume} {89}},\ \bibinfo
			{pages} {276601} (\bibinfo {year} {2002})}\BibitemShut {NoStop}%
		\bibitem [{\citenamefont {L{\'{o}}pez~Ortega}\ \emph
			{et~al.}(2015)\citenamefont {L{\'{o}}pez~Ortega}, \citenamefont {Estrader},
			\citenamefont {Salazar~Alvarez}, \citenamefont {Roca},\ and\ \citenamefont
			{Nogu{\'{e}}s}}]{LopezOrtega2015}%
		\BibitemOpen
		\bibfield  {author} {\bibinfo {author} {\bibfnamefont {Alberto}\ \bibnamefont
				{L{\'{o}}pez~Ortega}}, \bibinfo {author} {\bibfnamefont {Marta}\ \bibnamefont
				{Estrader}}, \bibinfo {author} {\bibfnamefont {German}\ \bibnamefont
				{Salazar~Alvarez}}, \bibinfo {author} {\bibfnamefont {Alejando~G.}\
				\bibnamefont {Roca}}, \ and\ \bibinfo {author} {\bibfnamefont {Josep}\
				\bibnamefont {Nogu{\'{e}}s}},\ }\bibfield  {title} {\enquote {\bibinfo
				{title} {Applications of exchange coupled bi$-$magnetic hard/soft and
					soft/hard magnetic core/shell nanoparticles},}\ }\href {\doibase
			10.1016/j.physrep.2014.09.007} {\bibfield  {journal} {\bibinfo  {journal}
				{Phys. Rep.}\ }\textbf {\bibinfo {volume} {553}},\ \bibinfo {pages} {1--32}
			(\bibinfo {year} {2015})}\BibitemShut {NoStop}%
		\bibitem [{\citenamefont {Lavorato}\ \emph {et~al.}(2017)\citenamefont
			{Lavorato}, \citenamefont {Lima}, \citenamefont {Troiani}, \citenamefont
			{Zysler},\ and\ \citenamefont {Winkler}}]{Lavorato2017}%
		\BibitemOpen
		\bibfield  {author} {\bibinfo {author} {\bibfnamefont {G.~C.}\ \bibnamefont
				{Lavorato}}, \bibinfo {author} {\bibfnamefont {E.}~\bibnamefont {Lima}},
			\bibinfo {author} {\bibfnamefont {H.~E.}\ \bibnamefont {Troiani}}, \bibinfo
			{author} {\bibfnamefont {R.~D.}\ \bibnamefont {Zysler}}, \ and\ \bibinfo
			{author} {\bibfnamefont {E.~L.}\ \bibnamefont {Winkler}},\ }\bibfield
		{title} {\enquote {\bibinfo {title} {Tuning the coercivity and exchange bias
					by controlling the interface coupling in bimagnetic core/shell
					nanoparticles},}\ }\href {\doibase 10.1039/C7NR03740F} {\bibfield  {journal}
			{\bibinfo  {journal} {Nanoscale}\ }\textbf {\bibinfo {volume} {9}},\ \bibinfo
			{pages} {10240--10247} (\bibinfo {year} {2017})}\BibitemShut {NoStop}%
		\bibitem [{\citenamefont {Salazar-Alvarez}\ \emph {et~al.}(2007)\citenamefont
			{Salazar-Alvarez}, \citenamefont {Sort}, \citenamefont {Suri{\~{n}}ach},
			\citenamefont {Bar{\'{o}}},\ and\ \citenamefont
			{Nogu{\'{e}}s}}]{Salazar-Alvarez2007}%
		\BibitemOpen
		\bibfield  {author} {\bibinfo {author} {\bibfnamefont {G.}~\bibnamefont
				{Salazar-Alvarez}}, \bibinfo {author} {\bibfnamefont {J.}~\bibnamefont
				{Sort}}, \bibinfo {author} {\bibfnamefont {S.}~\bibnamefont
				{Suri{\~{n}}ach}}, \bibinfo {author} {\bibfnamefont {M.~D.}\ \bibnamefont
				{Bar{\'{o}}}}, \ and\ \bibinfo {author} {\bibfnamefont {J.}~\bibnamefont
				{Nogu{\'{e}}s}},\ }\bibfield  {title} {\enquote {\bibinfo {title} {{Synthesis
						and Size-Dependent Exchange Bias in Inverted Core/Shell MnO/Mn$_3$O$_4$
						Nanoparticles}},}\ }\href {\doibase 10.1021/ja0714282} {\bibfield  {journal}
			{\bibinfo  {journal} {{J. Am. Chem. Soc.}}\ }\textbf {\bibinfo {volume}
				{129}},\ \bibinfo {pages} {9102--9108} (\bibinfo {year} {2007})}\BibitemShut
		{NoStop}%
		\bibitem [{\citenamefont {Skumryev}\ \emph {et~al.}(2003)\citenamefont
			{Skumryev}, \citenamefont {Stoyanov}, \citenamefont {Zhang}, \citenamefont
			{Hadjipanayis}, \citenamefont {Givord},\ and\ \citenamefont
			{Nogu{\'{e}}s}}]{Skumryev2003}%
		\BibitemOpen
		\bibfield  {author} {\bibinfo {author} {\bibfnamefont {V.}~\bibnamefont
				{Skumryev}}, \bibinfo {author} {\bibfnamefont {S.}~\bibnamefont {Stoyanov}},
			\bibinfo {author} {\bibfnamefont {Y.}~\bibnamefont {Zhang}}, \bibinfo
			{author} {\bibfnamefont {G.}~\bibnamefont {Hadjipanayis}}, \bibinfo {author}
			{\bibfnamefont {D.}~\bibnamefont {Givord}}, \ and\ \bibinfo {author}
			{\bibfnamefont {J.}~\bibnamefont {Nogu{\'{e}}s}},\ }\bibfield  {title}
		{\enquote {\bibinfo {title} {Beating the superparamagnetic limit with
					exchange bias},}\ }\href {\doibase 10.1038/nature01687} {\bibfield  {journal}
			{\bibinfo  {journal} {Nature}\ }\textbf {\bibinfo {volume} {423}},\ \bibinfo
			{pages} {850--853} (\bibinfo {year} {2003})}\BibitemShut {NoStop}%
		\bibitem [{\citenamefont {Sarkar}\ \emph {et~al.}(2012)\citenamefont {Sarkar},
			\citenamefont {Behera}, \citenamefont {Adhikari},\ and\ \citenamefont
			{Das}}]{Sarkar2012}%
		\BibitemOpen
		\bibfield  {author} {\bibinfo {author} {\bibfnamefont {A.}~\bibnamefont
				{Sarkar}}, \bibinfo {author} {\bibfnamefont {N.}~\bibnamefont {Behera}},
			\bibinfo {author} {\bibfnamefont {R.}~\bibnamefont {Adhikari}}, \ and\
			\bibinfo {author} {\bibfnamefont {A.~K.}\ \bibnamefont {Das}},\ }\bibfield
		{title} {\enquote {\bibinfo {title} {Studies on nonlinear electrical
					transport and magnetoresistance in co/coo core-shell nanostructure},}\ }\href
		{\doibase 10.1063/1.4710311} {\bibfield  {journal} {\bibinfo  {journal} {AIP
					Conference Proceedings}\ }\textbf {\bibinfo {volume} {1447}},\ \bibinfo
			{pages} {937--938} (\bibinfo {year} {2012})}\BibitemShut {NoStop}%
		\bibitem [{\citenamefont {Winkler}\ \emph {et~al.}(2012)\citenamefont
			{Winkler}, \citenamefont {Lima~Jr.}, \citenamefont {Tobia}, \citenamefont
			{Saleta}, \citenamefont {Troiani}, \citenamefont {Agostinelli}, \citenamefont
			{Fiorani},\ and\ \citenamefont {Zysler}}]{Winkler2012}%
		\BibitemOpen
		\bibfield  {author} {\bibinfo {author} {\bibfnamefont {E.~L.}\ \bibnamefont
				{Winkler}}, \bibinfo {author} {\bibfnamefont {E.}~\bibnamefont {Lima~Jr.}},
			\bibinfo {author} {\bibfnamefont {D.}~\bibnamefont {Tobia}}, \bibinfo
			{author} {\bibfnamefont {M.~E.}\ \bibnamefont {Saleta}}, \bibinfo {author}
			{\bibfnamefont {H.~E.}\ \bibnamefont {Troiani}}, \bibinfo {author}
			{\bibfnamefont {E.}~\bibnamefont {Agostinelli}}, \bibinfo {author}
			{\bibfnamefont {D.}~\bibnamefont {Fiorani}}, \ and\ \bibinfo {author}
			{\bibfnamefont {R.}~\bibnamefont {Zysler}},\ }\bibfield  {title} {\enquote
			{\bibinfo {title} {Origin of magnetic anisotropy in {ZnO/CoFe$_2$O$_4$} and
					{CoO/CoFe$_2$O$_4$} core/shell nanoparticle systems},}\ }\href {\doibase
			10.1063/1.4771993} {\bibfield  {journal} {\bibinfo  {journal} {Appl. Phys.
					Lett.}\ }\textbf {\bibinfo {volume} {101}},\ \bibinfo {pages} {252405}
			(\bibinfo {year} {2012})}\BibitemShut {NoStop}%
		\bibitem [{\citenamefont {Lavorato}\ \emph
			{et~al.}(2016{\natexlab{a}})\citenamefont {Lavorato}, \citenamefont
			{Winkler}, \citenamefont {Ghirri}, \citenamefont {Lima~Jr}, \citenamefont
			{Troiani}, \citenamefont {Fiorani}, \citenamefont {Agostinelli},
			\citenamefont {Rinaldi},\ and\ \citenamefont {Zysler}}]{Lavorato2016}%
		\BibitemOpen
		\bibfield  {author} {\bibinfo {author} {\bibfnamefont {G.}~\bibnamefont
				{Lavorato}}, \bibinfo {author} {\bibfnamefont {E.}~\bibnamefont {Winkler}},
			\bibinfo {author} {\bibfnamefont {A.}~\bibnamefont {Ghirri}}, \bibinfo
			{author} {\bibfnamefont {D.}~\bibnamefont {Lima~Jr}, \bibfnamefont
				{E.and~Peddis}}, \bibinfo {author} {\bibfnamefont {H.~E}\ \bibnamefont
				{Troiani}}, \bibinfo {author} {\bibfnamefont {D.}~\bibnamefont {Fiorani}},
			\bibinfo {author} {\bibfnamefont {E.}~\bibnamefont {Agostinelli}}, \bibinfo
			{author} {\bibfnamefont {D.}~\bibnamefont {Rinaldi}}, \ and\ \bibinfo
			{author} {\bibfnamefont {R.~D.}\ \bibnamefont {Zysler}},\ }\bibfield  {title}
		{\enquote {\bibinfo {title} {Exchange bias and surface effects in bimagnetic
					{CoO-Core/Co$_{0.5}$Ni$_{0.5}$Fe$_2$O$_4$-Shell} nanoparticles},}\ }\href
		{\doibase 10.1063/1.4771993} {\bibfield  {journal} {\bibinfo  {journal}
				{Physical Review B}\ }\textbf {\bibinfo {volume} {94}},\ \bibinfo {pages}
			{054432} (\bibinfo {year} {2016}{\natexlab{a}})}\BibitemShut {NoStop}%
		\bibitem [{\citenamefont {Liu}\ \emph {et~al.}(2015)\citenamefont {Liu},
			\citenamefont {Pichon}, \citenamefont {Ulhaq}, \citenamefont {Lef{\`{e}}vre},
			\citenamefont {Gren{\`{e}}che}, \citenamefont {B{\'{e}}gin},\ and\
			\citenamefont {B{\'{e}}gin-Colin}}]{Liu2015}%
		\BibitemOpen
		\bibfield  {author} {\bibinfo {author} {\bibfnamefont {X.}~\bibnamefont
				{Liu}}, \bibinfo {author} {\bibfnamefont {B.~P.}\ \bibnamefont {Pichon}},
			\bibinfo {author} {\bibfnamefont {C.}~\bibnamefont {Ulhaq}}, \bibinfo
			{author} {\bibfnamefont {C.}~\bibnamefont {Lef{\`{e}}vre}}, \bibinfo {author}
			{\bibfnamefont {J.~M.}\ \bibnamefont {Gren{\`{e}}che}}, \bibinfo {author}
			{\bibfnamefont {D.}~\bibnamefont {B{\'{e}}gin}}, \ and\ \bibinfo {author}
			{\bibfnamefont {S.}~\bibnamefont {B{\'{e}}gin-Colin}},\ }\bibfield  {title}
		{\enquote {\bibinfo {title} {{Systematic Study of Exchange Coupling in
						Core-Shell Fe $_{3-delta}$O$_4/$CoO Nanoparticles}},}\ }\href {\doibase
			10.1021/acs.chemmater.5b01103} {\bibfield  {journal} {\bibinfo  {journal}
				{Chemistry of Materials}\ }\textbf {\bibinfo {volume} {27}},\ \bibinfo
			{pages} {4073--4081} (\bibinfo {year} {2015})},\ \Eprint
		{http://arxiv.org/abs/https://doi.org/10.1021/acs.chemmater.5b01103}
		{https://doi.org/10.1021/acs.chemmater.5b01103} \BibitemShut {NoStop}%
		\bibitem [{\citenamefont {{Lottini, E. and L{\'o}pez-Ortega, A. and Bertoni, G.
					and Turner, S. and Meledina, M. and Tendeloo, G. Van and de Juli{\'a}n
					Fern{\'a}ndez, C. and Sangregorio, C.}}(2016)}]{Lottini2016}%
		\BibitemOpen
		\bibfield  {author} {\bibinfo {author} {\bibnamefont {{Lottini, E. and
						L{\'o}pez-Ortega, A. and Bertoni, G. and Turner, S. and Meledina, M. and
						Tendeloo, G. Van and de Juli{\'a}n Fern{\'a}ndez, C. and Sangregorio, C.}}},\
		}\bibfield  {title} {\enquote {\bibinfo {title} {{Strongly Exchange Coupled
						Core/Shell Nanoparticles with High Magnetic Anisotropy:A Strategy Toward
						Rare-Earth-Free Permanent Magnets}},}\ }\href {\doibase
			10.1021/acs.chemmater.6b00623} {\bibfield  {journal} {\bibinfo  {journal}
				{{Chemistry of Materials}}\ }\textbf {\bibinfo {volume} {28}},\ \bibinfo
			{pages} {4214--4222} (\bibinfo {year} {2016})}\BibitemShut {NoStop}%
		\bibitem [{\citenamefont {Lavorato}\ \emph {et~al.}(2014)\citenamefont
			{Lavorato}, \citenamefont {{Lima Jr}}, \citenamefont {Tobia}, \citenamefont
			{Fiorani}, \citenamefont {Troiani}, \citenamefont {Zysler},\ and\
			\citenamefont {~}}]{Lavorato2014}%
		\BibitemOpen
		\bibfield  {author} {\bibinfo {author} {\bibfnamefont {Gabriel~C}\
				\bibnamefont {Lavorato}}, \bibinfo {author} {\bibfnamefont {Enio}\
				\bibnamefont {{Lima Jr}}}, \bibinfo {author} {\bibfnamefont {Dina}\
				\bibnamefont {Tobia}}, \bibinfo {author} {\bibfnamefont {Dino}\ \bibnamefont
				{Fiorani}}, \bibinfo {author} {\bibfnamefont {Horacio~E}\ \bibnamefont
				{Troiani}}, \bibinfo {author} {\bibfnamefont {Roberto~D}\ \bibnamefont
				{Zysler}}, \ and\ \bibinfo {author} {\bibfnamefont {Elin L.~Winkler}\
				\bibnamefont {~}},\ }\bibfield  {title} {\enquote {\bibinfo {title} {{Size
						Effects in Bimagnetic CoO/CoFe$_2$O$_4$ Core/Shell Nanoparticles.}}}\ }\href
		{\doibase 10.1088/0957-4484/25/35/355704} {\bibfield  {journal} {\bibinfo
				{journal} {{Nanotechnology}}\ }\textbf {\bibinfo {volume} {25}},\ \bibinfo
			{pages} {355704} (\bibinfo {year} {2014})}\BibitemShut {NoStop}%
		\bibitem [{\citenamefont {Sytnyk}\ \emph {et~al.}(2013)\citenamefont {Sytnyk},
			\citenamefont {Kirchschlager}, \citenamefont {Bodnarchuk}, \citenamefont
			{Primetzhofer}, \citenamefont {Kriegner}, \citenamefont {Enser},
			\citenamefont {Stangl}, \citenamefont {Bauer}, \citenamefont {Voith},
			\citenamefont {Hassel}, \citenamefont {Krumeich}, \citenamefont {Ludwig},
			\citenamefont {Meingast}, \citenamefont {Kothleitner}, \citenamefont
			{Kovalenko},\ and\ \citenamefont {Heiss}}]{Sytnyk2013}%
		\BibitemOpen
		\bibfield  {author} {\bibinfo {author} {\bibfnamefont {M.}~\bibnamefont
				{Sytnyk}}, \bibinfo {author} {\bibfnamefont {R.}~\bibnamefont
				{Kirchschlager}}, \bibinfo {author} {\bibfnamefont {M.~I.}\ \bibnamefont
				{Bodnarchuk}}, \bibinfo {author} {\bibfnamefont {D.}~\bibnamefont
				{Primetzhofer}}, \bibinfo {author} {\bibfnamefont {D.}~\bibnamefont
				{Kriegner}}, \bibinfo {author} {\bibfnamefont {H.}~\bibnamefont {Enser}},
			\bibinfo {author} {\bibfnamefont {J.}~\bibnamefont {Stangl}}, \bibinfo
			{author} {\bibfnamefont {P.}~\bibnamefont {Bauer}}, \bibinfo {author}
			{\bibfnamefont {M.}~\bibnamefont {Voith}}, \bibinfo {author} {\bibfnamefont
				{A.~W.}\ \bibnamefont {Hassel}}, \bibinfo {author} {\bibfnamefont
				{F.}~\bibnamefont {Krumeich}}, \bibinfo {author} {\bibfnamefont
				{F.}~\bibnamefont {Ludwig}}, \bibinfo {author} {\bibfnamefont
				{A.}~\bibnamefont {Meingast}}, \bibinfo {author} {\bibfnamefont
				{G.}~\bibnamefont {Kothleitner}}, \bibinfo {author} {\bibfnamefont
				{M.}~\bibnamefont {Kovalenko}}, \ and\ \bibinfo {author} {\bibfnamefont
				{W.}~\bibnamefont {Heiss}},\ }\bibfield  {title} {\enquote {\bibinfo {title}
				{Tuning the magnetic properties of metal oxide nanocrystal heterostructures
					by cation exchange},}\ }\href {\doibase 10.1021/nl304115r} {\bibfield
			{journal} {\bibinfo  {journal} {{Nano Letters}}\ }\textbf {\bibinfo {volume}
				{13}},\ \bibinfo {pages} {586--593} (\bibinfo {year} {2013})},\ \bibinfo
		{note} {pMID: 23362940}\BibitemShut {NoStop}%
		\bibitem [{\citenamefont {Fabris}\ \emph {et~al.}(2019)\citenamefont {Fabris},
			\citenamefont {Lima~Jr.}, \citenamefont {De~Biasi}, \citenamefont {Troiani},
			\citenamefont {Vasquez~Mansilla}, \citenamefont {Torres}, \citenamefont
			{Fern\'{a}ndez-Pacheco}, \citenamefont {Ibarra}, \citenamefont {Goya},
			\citenamefont {R.},\ and\ \citenamefont {Winkler}}]{Fabris2019}%
		\BibitemOpen
		\bibfield  {author} {\bibinfo {author} {\bibfnamefont {F.}~\bibnamefont
				{Fabris}}, \bibinfo {author} {\bibfnamefont {E.}~\bibnamefont {Lima~Jr.}},
			\bibinfo {author} {\bibfnamefont {E.}~\bibnamefont {De~Biasi}}, \bibinfo
			{author} {\bibfnamefont {H.~E.}\ \bibnamefont {Troiani}}, \bibinfo {author}
			{\bibfnamefont {M.}~\bibnamefont {Vasquez~Mansilla}}, \bibinfo {author}
			{\bibfnamefont {T.~E.}\ \bibnamefont {Torres}}, \bibinfo {author}
			{\bibfnamefont {R.}~\bibnamefont {Fern\'{a}ndez-Pacheco}}, \bibinfo {author}
			{\bibfnamefont {M.~R.}\ \bibnamefont {Ibarra}}, \bibinfo {author}
			{\bibfnamefont {G.~F.}\ \bibnamefont {Goya}}, \bibinfo {author}
			{\bibfnamefont {Zysler}\ \bibnamefont {R.}}, \ and\ \bibinfo {author}
			{\bibfnamefont {E.}~\bibnamefont {Winkler}},\ }\bibfield  {title} {\enquote
			{\bibinfo {title} {Controlling the dominant magnetic relaxation mechanisms
					for magnetic hyperthermia in bimagnetic core-shell nanoparticles},}\ }\href
		{\doibase 10.1039/C8NR07834C} {\bibfield  {journal} {\bibinfo  {journal}
				{Nanoscale}\ }\textbf {\bibinfo {volume} {11}},\ \bibinfo {pages}
			{3164--3172} (\bibinfo {year} {2019})}\BibitemShut {NoStop}%
		\bibitem [{\citenamefont {Lavorato}\ \emph {et~al.}(2015)\citenamefont
			{Lavorato}, \citenamefont {Lima}, \citenamefont {Troiani}, \citenamefont
			{Zysler},\ and\ \citenamefont {Winkler}}]{Lavorato2015}%
		\BibitemOpen
		\bibfield  {author} {\bibinfo {author} {\bibfnamefont {G.C.}\ \bibnamefont
				{Lavorato}}, \bibinfo {author} {\bibfnamefont {E.}~\bibnamefont {Lima}},
			\bibinfo {author} {\bibfnamefont {H.E.}\ \bibnamefont {Troiani}}, \bibinfo
			{author} {\bibfnamefont {R.D.}\ \bibnamefont {Zysler}}, \ and\ \bibinfo
			{author} {\bibfnamefont {E.L.}\ \bibnamefont {Winkler}},\ }\bibfield  {title}
		{\enquote {\bibinfo {title} {Exchange-coupling in thermal annealed bimagnetic
					core/shell nanoparticles},}\ }\href {\doibase
			https://doi.org/10.1016/j.jallcom.2015.02.050} {\bibfield  {journal}
			{\bibinfo  {journal} {Journal of Alloys and Compounds}\ }\textbf {\bibinfo
				{volume} {633}},\ \bibinfo {pages} {333 -- 337} (\bibinfo {year}
			{2015})}\BibitemShut {NoStop}%
		\bibitem [{\citenamefont {Sun}\ and\ \citenamefont {Zeng}(2002)}]{Sun2002}%
		\BibitemOpen
		\bibfield  {author} {\bibinfo {author} {\bibfnamefont {S.}~\bibnamefont
				{Sun}}\ and\ \bibinfo {author} {\bibfnamefont {H.}~\bibnamefont {Zeng}},\
		}\bibfield  {title} {\enquote {\bibinfo {title} {Size-controlled synthesis of
					magnetite nanoparticles},}\ }\href {\doibase 10.1021/ja026501x} {\bibfield
			{journal} {\bibinfo  {journal} {Journal of the American Chemical Society}\
			}\textbf {\bibinfo {volume} {124}},\ \bibinfo {pages} {8204--8205} (\bibinfo
			{year} {2002})},\ \bibinfo {note} {pMID: 12105897}\BibitemShut {NoStop}%
		\bibitem [{\citenamefont {Sun}\ \emph {et~al.}(2004)\citenamefont {Sun},
			\citenamefont {Zeng}, \citenamefont {Robinson}, \citenamefont {Raoux},
			\citenamefont {Rice}, \citenamefont {Wang},\ and\ \citenamefont
			{Li}}]{Sun2004}%
		\BibitemOpen
		\bibfield  {author} {\bibinfo {author} {\bibfnamefont {S.~H}\ \bibnamefont
				{Sun}}, \bibinfo {author} {\bibfnamefont {H.}~\bibnamefont {Zeng}}, \bibinfo
			{author} {\bibfnamefont {D.~B.}\ \bibnamefont {Robinson}}, \bibinfo {author}
			{\bibfnamefont {S.}~\bibnamefont {Raoux}}, \bibinfo {author} {\bibfnamefont
				{P.~M.}\ \bibnamefont {Rice}}, \bibinfo {author} {\bibfnamefont {S.~X.}\
				\bibnamefont {Wang}}, \ and\ \bibinfo {author} {\bibfnamefont {G.~X.}\
				\bibnamefont {Li}},\ }\bibfield  {title} {\enquote {\bibinfo {title}
				{Monodisperse {MFe$_{2}$O$_{4}$} ({M = Fe, Co, Mn}) nanoparticles},}\ }\href
		{\doibase 10.1021/ja0380852} {\bibfield  {journal} {\bibinfo  {journal}
				{Journal of the American Chemical Society}\ }\textbf {\bibinfo {volume}
				{126}},\ \bibinfo {pages} {273--279} (\bibinfo {year} {2004})}\BibitemShut
		{NoStop}%
		\bibitem [{Sup()}]{SupMat}%
		\BibitemOpen
		\bibfield  {title} {\enquote {\bibinfo {title} {{See Supplemental Material at
						URL for additional microstructural information of the core/shell
						nanoparticles assemblies annealed at different temperature, obtained by TEM
						and HRTEM microscopy. Further characterization of the as-synthetized and
						annealed Fe$_{3}$O$_{4}$/CoFe$_{2}$O$_{4}$ core/shell nanoparticles by
						thermogravimetric (TGA) and infrared (FT-IR) measurements, and additional
						magnetic characterization, as the hysteresis cycles as a function of the
						temperature and ZFC and FC magnetization curves of the annealed
						nanoparticles, it is also included. See also details of the magnetoresistance
						temperature evolution of the self-assemblies.}}}\ }\href@noop {} {\
		}\BibitemShut {NoStop}%
		\bibitem [{\citenamefont {L\'{o}pez-Ortega}\ \emph {et~al.}(2012)\citenamefont
			{L\'{o}pez-Ortega}, \citenamefont {Estrader}, \citenamefont
			{Salazar-Alvarez}, \citenamefont {Estrad\'{e}}, \citenamefont {Golosovsky},
			\citenamefont {Dumas}, \citenamefont {Keavney}, \citenamefont {Vasilakaki},
			\citenamefont {Trohidou}, \citenamefont {Sort}, \citenamefont {Peir\'{o}},
			\citenamefont {Suri\~{n}ach}, \citenamefont {Bar\'{o}},\ and\ \citenamefont
			{Nogu\'{e}s}}]{Lopez2012}%
		\BibitemOpen
		\bibfield  {author} {\bibinfo {author} {\bibfnamefont {A.}~\bibnamefont
				{L\'{o}pez-Ortega}}, \bibinfo {author} {\bibfnamefont {M.}~\bibnamefont
				{Estrader}}, \bibinfo {author} {\bibfnamefont {G.}~\bibnamefont
				{Salazar-Alvarez}}, \bibinfo {author} {\bibfnamefont {S.}~\bibnamefont
				{Estrad\'{e}}}, \bibinfo {author} {\bibfnamefont {I.~V.}\ \bibnamefont
				{Golosovsky}}, \bibinfo {author} {\bibfnamefont {R.~K.}\ \bibnamefont
				{Dumas}}, \bibinfo {author} {\bibfnamefont {D.~J.}\ \bibnamefont {Keavney}},
			\bibinfo {author} {\bibfnamefont {M.}~\bibnamefont {Vasilakaki}}, \bibinfo
			{author} {\bibfnamefont {K.~N.}\ \bibnamefont {Trohidou}}, \bibinfo {author}
			{\bibfnamefont {J.}~\bibnamefont {Sort}}, \bibinfo {author} {\bibfnamefont
				{F.}~\bibnamefont {Peir\'{o}}}, \bibinfo {author} {\bibfnamefont
				{S.}~\bibnamefont {Suri\~{n}ach}}, \bibinfo {author} {\bibfnamefont {M.~D.}\
				\bibnamefont {Bar\'{o}}}, \ and\ \bibinfo {author} {\bibfnamefont
				{J.}~\bibnamefont {Nogu\'{e}s}},\ }\bibfield  {title} {\enquote {\bibinfo
				{title} {Strongly exchange coupled inverse ferrimagnetic soft/hard
					{Mn$_{x}$Fe$_{3 - x}$O$_{4}/$Fe$_{x}$Mn$_{3 -x}$O$_{4}$} core/shell
					heterostructured nanoparticles},}\ }\href {\doibase 10.1039/C2NR30986F}
		{\bibfield  {journal} {\bibinfo  {journal} {Nanoscale}\ }\textbf {\bibinfo
				{volume} {4}},\ \bibinfo {pages} {5138--5147} (\bibinfo {year}
			{2012})}\BibitemShut {NoStop}%
		\bibitem [{\citenamefont {Krycka}\ \emph {et~al.}(2013)\citenamefont {Krycka},
			\citenamefont {Borchers}, \citenamefont {Salazar-Alvarez}, \citenamefont
			{L\'{o}pez-Ortega}, \citenamefont {Estrader}, \citenamefont {Estrad\'{e}},
			\citenamefont {Winkler}, \citenamefont {Zysler}, \citenamefont {Sort},
			\citenamefont {Peir\'{o}}, \citenamefont {Bar\'{o}}, \citenamefont {Kao},\
			and\ \citenamefont {Nogu\'{e}s}}]{Krycka2013}%
		\BibitemOpen
		\bibfield  {author} {\bibinfo {author} {\bibfnamefont {K.~L.}\ \bibnamefont
				{Krycka}}, \bibinfo {author} {\bibfnamefont {J.~A.}\ \bibnamefont
				{Borchers}}, \bibinfo {author} {\bibfnamefont {G.}~\bibnamefont
				{Salazar-Alvarez}}, \bibinfo {author} {\bibfnamefont {A.}~\bibnamefont
				{L\'{o}pez-Ortega}}, \bibinfo {author} {\bibfnamefont {M.}~\bibnamefont
				{Estrader}}, \bibinfo {author} {\bibfnamefont {S.}~\bibnamefont
				{Estrad\'{e}}}, \bibinfo {author} {\bibfnamefont {E.}~\bibnamefont
				{Winkler}}, \bibinfo {author} {\bibfnamefont {R.~D.}\ \bibnamefont {Zysler}},
			\bibinfo {author} {\bibfnamefont {J.}~\bibnamefont {Sort}}, \bibinfo {author}
			{\bibfnamefont {F.}~\bibnamefont {Peir\'{o}}}, \bibinfo {author}
			{\bibfnamefont {M.~D.}\ \bibnamefont {Bar\'{o}}}, \bibinfo {author}
			{\bibfnamefont {C.C.}\ \bibnamefont {Kao}}, \ and\ \bibinfo {author}
			{\bibfnamefont {J.}~\bibnamefont {Nogu\'{e}s}},\ }\bibfield  {title}
		{\enquote {\bibinfo {title} {Resolving material-specific structures within
					{Fe3O4/$\gamma$-Mn2O3} core/shell nanoparticles using anomalous small-angle
					x-ray scattering},}\ }\href {\doibase 10.1021/nn303600e} {\bibfield
			{journal} {\bibinfo  {journal} {ACS Nano}\ }\textbf {\bibinfo {volume} {7}},\
			\bibinfo {pages} {921--931} (\bibinfo {year} {2013})},\ \bibinfo {note}
		{pMID: 23320459}\BibitemShut {NoStop}%
		\bibitem [{\citenamefont {Skomski}\ and\ \citenamefont
			{Coey}(1993)}]{Skomsky1993}%
		\BibitemOpen
		\bibfield  {author} {\bibinfo {author} {\bibfnamefont {R.}~\bibnamefont
				{Skomski}}\ and\ \bibinfo {author} {\bibfnamefont {J.~M.~D.}\ \bibnamefont
				{Coey}},\ }\bibfield  {title} {\enquote {\bibinfo {title} {Giant energy
					product in nanostructured two-phase magnets},}\ }\href {\doibase
			10.1103/PhysRevB.48.15812} {\bibfield  {journal} {\bibinfo  {journal} {Phys.
					Rev. B}\ }\textbf {\bibinfo {volume} {48}},\ \bibinfo {pages} {15812--15816}
			(\bibinfo {year} {1993})}\BibitemShut {NoStop}%
		\bibitem [{\citenamefont {Kneller}\ and\ \citenamefont
			{Hawig}(1991)}]{Kneller1991}%
		\BibitemOpen
		\bibfield  {author} {\bibinfo {author} {\bibfnamefont {E.~F.}\ \bibnamefont
				{Kneller}}\ and\ \bibinfo {author} {\bibfnamefont {R.}~\bibnamefont
				{Hawig}},\ }\bibfield  {title} {\enquote {\bibinfo {title} {{The
						Exchange-Spring Magnet: A New Material Principle for Permanent Magnets}},}\
		}\href {\doibase 10.1109/20.102931} {\bibfield  {journal} {\bibinfo
				{journal} {{IEEE Trans. Magn.}}\ }\textbf {\bibinfo {volume} {27}},\ \bibinfo
			{pages} {3588--3600} (\bibinfo {year} {1991})}\BibitemShut {NoStop}%
		\bibitem [{\citenamefont {Zhao}\ and\ \citenamefont {Wang}(2006)}]{Zhao2006}%
		\BibitemOpen
		\bibfield  {author} {\bibinfo {author} {\bibfnamefont {G.~P.}\ \bibnamefont
				{Zhao}}\ and\ \bibinfo {author} {\bibfnamefont {X.~L.}\ \bibnamefont
				{Wang}},\ }\bibfield  {title} {\enquote {\bibinfo {title} {Nucleation,
					pinning, and coercivity in magnetic nanosystems: An analytical micromagnetic
					approach},}\ }\href {\doibase 10.1103/PhysRevB.74.012409} {\bibfield
			{journal} {\bibinfo  {journal} {Phys. Rev. B}\ }\textbf {\bibinfo {volume}
				{74}},\ \bibinfo {pages} {012409} (\bibinfo {year} {2006})}\BibitemShut
		{NoStop}%
		\bibitem [{\citenamefont {Zhao}\ \emph {et~al.}(2005)\citenamefont {Zhao},
			\citenamefont {Zhao}, \citenamefont {Lim}, \citenamefont {Feng},\ and\
			\citenamefont {Ong}}]{Zhao2005}%
		\BibitemOpen
		\bibfield  {author} {\bibinfo {author} {\bibfnamefont {G.~P.}\ \bibnamefont
				{Zhao}}, \bibinfo {author} {\bibfnamefont {M.~G.}\ \bibnamefont {Zhao}},
			\bibinfo {author} {\bibfnamefont {H.~S.}\ \bibnamefont {Lim}}, \bibinfo
			{author} {\bibfnamefont {Y.~P.}\ \bibnamefont {Feng}}, \ and\ \bibinfo
			{author} {\bibfnamefont {C.~K.}\ \bibnamefont {Ong}},\ }\bibfield  {title}
		{\enquote {\bibinfo {title} {From nucleation to coercivity},}\ }\href
		{\doibase 10.1063/1.2108120} {\bibfield  {journal} {\bibinfo  {journal}
				{Applied Physics Letters}\ }\textbf {\bibinfo {volume} {87}},\ \bibinfo
			{pages} {162513} (\bibinfo {year} {2005})}\BibitemShut {NoStop}%
		\bibitem [{\citenamefont {Leineweber}\ and\ \citenamefont
			{Kronmüller}(1997)}]{Leineweber1997}%
		\BibitemOpen
		\bibfield  {author} {\bibinfo {author} {\bibfnamefont {T.}~\bibnamefont
				{Leineweber}}\ and\ \bibinfo {author} {\bibfnamefont {H.}~\bibnamefont
				{Kronmüller}},\ }\bibfield  {title} {\enquote {\bibinfo {title}
				{Micromagnetic examination of exchange coupled ferromagnetic nanolayers},}\
		}\href {\doibase https://doi.org/10.1016/S0304-8853(97)00601-X} {\bibfield
			{journal} {\bibinfo  {journal} {Journal of Magnetism and Magnetic Materials}\
			}\textbf {\bibinfo {volume} {176}},\ \bibinfo {pages} {145 -- 154} (\bibinfo
			{year} {1997})}\BibitemShut {NoStop}%
		\bibitem [{\citenamefont {Fullerton}\ \emph {et~al.}(1999)\citenamefont
			{Fullerton}, \citenamefont {Jiang},\ and\ \citenamefont
			{Bader}}]{Fullerton1999}%
		\BibitemOpen
		\bibfield  {author} {\bibinfo {author} {\bibfnamefont {Eric~E.}\ \bibnamefont
				{Fullerton}}, \bibinfo {author} {\bibfnamefont {J.~S.}\ \bibnamefont
				{Jiang}}, \ and\ \bibinfo {author} {\bibfnamefont {S.~D.}\ \bibnamefont
				{Bader}},\ }\bibfield  {title} {\enquote {\bibinfo {title} {{Hard/Soft
						Magnetic Heterostructures: Model Exchange-Spring Magnets}},}\ }\href
		{\doibase 10.1016/S0304-8853(99)00376-5} {\bibfield  {journal} {\bibinfo
				{journal} {J. Magn. Magn. Mater.}\ }\textbf {\bibinfo {volume} {200}},\
			\bibinfo {pages} {392--404} (\bibinfo {year} {1999})}\BibitemShut {NoStop}%
		\bibitem [{\citenamefont {Zhao}\ \emph {et~al.}(2007)\citenamefont {Zhao},
			\citenamefont {Wang}, \citenamefont {Feng},\ and\ \citenamefont
			{Huang}}]{Zhao2007b}%
		\BibitemOpen
		\bibfield  {author} {\bibinfo {author} {\bibfnamefont {G.~P.}\ \bibnamefont
				{Zhao}}, \bibinfo {author} {\bibfnamefont {X.~L.}\ \bibnamefont {Wang}},
			\bibinfo {author} {\bibfnamefont {Y.~P.}\ \bibnamefont {Feng}}, \ and\
			\bibinfo {author} {\bibfnamefont {C.~W.}\ \bibnamefont {Huang}},\ }\bibfield
		{title} {\enquote {\bibinfo {title} {Coherent rotation and effective
					anisotropy},}\ }\href {\doibase 10.1109/TMAG.2007.893629} {\bibfield
			{journal} {\bibinfo  {journal} {IEEE Transactions on Magnetics}\ }\textbf
			{\bibinfo {volume} {43}},\ \bibinfo {pages} {2908--2910} (\bibinfo {year}
			{2007})}\BibitemShut {NoStop}%
		\bibitem [{\citenamefont {Coey}(2010)}]{Coey2010}%
		\BibitemOpen
		\bibfield  {author} {\bibinfo {author} {\bibfnamefont {J.~M.~D.}\
				\bibnamefont {Coey}},\ }\href {\doibase 10.1017/CBO9780511845000} {\emph
			{\bibinfo {title} {Magnetism and Magnetic Materials}}}\ (\bibinfo
		{publisher} {Cambridge University Press},\ \bibinfo {year}
		{2010})\BibitemShut {NoStop}%
		\bibitem [{\citenamefont {Lavorato}\ \emph
			{et~al.}(2016{\natexlab{b}})\citenamefont {Lavorato}, \citenamefont
			{Winkler}, \citenamefont {Rivas-Murias},\ and\ \citenamefont
			{Rivadulla}}]{Lavorato2016b}%
		\BibitemOpen
		\bibfield  {author} {\bibinfo {author} {\bibfnamefont {G.}~\bibnamefont
				{Lavorato}}, \bibinfo {author} {\bibfnamefont {E.}~\bibnamefont {Winkler}},
			\bibinfo {author} {\bibfnamefont {B.}~\bibnamefont {Rivas-Murias}}, \ and\
			\bibinfo {author} {\bibfnamefont {F.}~\bibnamefont {Rivadulla}},\ }\bibfield
		{title} {\enquote {\bibinfo {title} {Thickness dependence of exchange
					coupling in epitaxial {Fe$_3$O$_4/$CoFe$_2$O$4$} soft/hard magnetic
					bilayers},}\ }\href {\doibase 10.1103/PhysRevB.94.054405} {\bibfield
			{journal} {\bibinfo  {journal} {Phys. Rev. B}\ }\textbf {\bibinfo {volume}
				{94}},\ \bibinfo {pages} {054405} (\bibinfo {year}
			{2016}{\natexlab{b}})}\BibitemShut {NoStop}%
		\bibitem [{\citenamefont {Bullita}\ \emph {et~al.}(2014)\citenamefont
			{Bullita}, \citenamefont {Casu}, \citenamefont {Casula}, \citenamefont
			{Concas}, \citenamefont {Congiu}, \citenamefont {Corrias}, \citenamefont
			{Falqui}, \citenamefont {Loche},\ and\ \citenamefont {Marras}}]{Bullita2014}%
		\BibitemOpen
		\bibfield  {author} {\bibinfo {author} {\bibfnamefont {S.}~\bibnamefont
				{Bullita}}, \bibinfo {author} {\bibfnamefont {A.}~\bibnamefont {Casu}},
			\bibinfo {author} {\bibfnamefont {M.~F.}\ \bibnamefont {Casula}}, \bibinfo
			{author} {\bibfnamefont {G.}~\bibnamefont {Concas}}, \bibinfo {author}
			{\bibfnamefont {F.}~\bibnamefont {Congiu}}, \bibinfo {author} {\bibfnamefont
				{A.}~\bibnamefont {Corrias}}, \bibinfo {author} {\bibfnamefont
				{A.}~\bibnamefont {Falqui}}, \bibinfo {author} {\bibfnamefont
				{D.}~\bibnamefont {Loche}}, \ and\ \bibinfo {author} {\bibfnamefont
				{C.}~\bibnamefont {Marras}},\ }\bibfield  {title} {\enquote {\bibinfo {title}
				{{ZnFe$_2$O$_4$} nanoparticles dispersed in a highly porous silica aerogel
					matrix: A magnetic study.}}\ }\href {\doibase 10.1039/c3cp54291b} {\bibfield
			{journal} {\bibinfo  {journal} {Phys. Chem. Chem. Phys.}\ }\textbf {\bibinfo
				{volume} {16}},\ \bibinfo {pages} {4843--52} (\bibinfo {year}
			{2014})}\BibitemShut {NoStop}%
		\bibitem [{\citenamefont {Dormann}\ \emph {et~al.}(1996)\citenamefont
			{Dormann}, \citenamefont {D'Orazio}, \citenamefont {Lucari}, \citenamefont
			{Tronc}, \citenamefont {Pren{\'{e}}}, \citenamefont {Jolivet}, \citenamefont
			{Fiorani}, \citenamefont {Cherkaoui},\ and\ \citenamefont
			{Nogu{\`{e}}s}}]{Dormann1996}%
		\BibitemOpen
		\bibfield  {author} {\bibinfo {author} {\bibfnamefont {J.}~\bibnamefont
				{Dormann}}, \bibinfo {author} {\bibfnamefont {F.}~\bibnamefont {D'Orazio}},
			\bibinfo {author} {\bibfnamefont {F.}~\bibnamefont {Lucari}}, \bibinfo
			{author} {\bibfnamefont {E.}~\bibnamefont {Tronc}}, \bibinfo {author}
			{\bibfnamefont {P.}~\bibnamefont {Pren{\'{e}}}}, \bibinfo {author}
			{\bibfnamefont {J.}~\bibnamefont {Jolivet}}, \bibinfo {author} {\bibfnamefont
				{D.}~\bibnamefont {Fiorani}}, \bibinfo {author} {\bibfnamefont
				{R.}~\bibnamefont {Cherkaoui}}, \ and\ \bibinfo {author} {\bibfnamefont
				{M.}~\bibnamefont {Nogu{\`{e}}s}},\ }\bibfield  {title} {\enquote {\bibinfo
				{title} {{Thermal Variation of the Relaxation Time of the Magnetic Moment of
						$\gamma$-Fe$_2$O$_3$ Nanoparticles with Interparticle Interactions of Various
						Strengths}},}\ }\href {\doibase 10.1103/PhysRevB.53.14291} {\bibfield
			{journal} {\bibinfo  {journal} {Phys. Rev. B}\ }\textbf {\bibinfo {volume}
				{53}},\ \bibinfo {pages} {14291--14297} (\bibinfo {year} {1996})}\BibitemShut
		{NoStop}%
		\bibitem [{\citenamefont {Simmons}(1963{\natexlab{a}})}]{Simmons1963}%
		\BibitemOpen
		\bibfield  {author} {\bibinfo {author} {\bibfnamefont {J.~G.}\ \bibnamefont
				{Simmons}},\ }\bibfield  {title} {\enquote {\bibinfo {title} {Generalized
					formula for the electric tunnel effect between similar electrodes separated
					by a thin insulating film},}\ }\href {https://doi.org/10.1063/1.1702682}
		{\bibfield  {journal} {\bibinfo  {journal} {Journal of Applied Physics}\
			}\textbf {\bibinfo {volume} {34}},\ \bibinfo {pages} {1793} (\bibinfo {year}
			{1963}{\natexlab{a}})}\BibitemShut {NoStop}%
		\bibitem [{\citenamefont {Simmons}(1963{\natexlab{b}})}]{Simmons1963_V3}%
		\BibitemOpen
		\bibfield  {author} {\bibinfo {author} {\bibfnamefont {J.~G.}\ \bibnamefont
				{Simmons}},\ }\bibfield  {title} {\enquote {\bibinfo {title} {Low-voltage
					current-voltage relationship of tunnel junctions},}\ }\href
		{https://doi.org/10.1063/1.1729081} {\bibfield  {journal} {\bibinfo
				{journal} {Journal of Applied Physics}\ }\textbf {\bibinfo {volume} {34}},\
			\bibinfo {pages} {238} (\bibinfo {year} {1963}{\natexlab{b}})}\BibitemShut
		{NoStop}%
		\bibitem [{\citenamefont {Vilan}(2007)}]{Vilan2007}%
		\BibitemOpen
		\bibfield  {author} {\bibinfo {author} {\bibfnamefont {A.}~\bibnamefont
				{Vilan}},\ }\bibfield  {title} {\enquote {\bibinfo {title} {Analyzing
					molecular current-voltage characteristics with the simmons tunneling model:
					Scaling and linearization},}\ }\href {\doibase 10.1021/jp066846s} {\bibfield
			{journal} {\bibinfo  {journal} {J. Phys. Chem. C}\ }\textbf {\bibinfo
				{volume} {111}},\ \bibinfo {pages} {4431--4444} (\bibinfo {year}
			{2007})}\BibitemShut {NoStop}%
		\bibitem [{\citenamefont {Efros}\ and\ \citenamefont
			{Shklovskii}(1975)}]{Efros1975}%
		\BibitemOpen
		\bibfield  {author} {\bibinfo {author} {\bibfnamefont {A.L.}\ \bibnamefont
				{Efros}}\ and\ \bibinfo {author} {\bibfnamefont {B.I.}\ \bibnamefont
				{Shklovskii}},\ }\bibfield  {title} {\enquote {\bibinfo {title} {Coulomb gap
					and low temperature conductivity of disordered systems},}\ }\href {\doibase
			10.1088/0022-3719/8/4/003} {\bibfield  {journal} {\bibinfo  {journal} {J.
					Phys. C: Solid State Phys.}\ }\textbf {\bibinfo {volume} {8}},\ \bibinfo
			{pages} {L49} (\bibinfo {year} {1975})}\BibitemShut {NoStop}%
		\bibitem [{\citenamefont {Song}\ \emph {et~al.}(2013)\citenamefont {Song},
			\citenamefont {Yang}, \citenamefont {Li}, \citenamefont {Li}, \citenamefont
			{Han}, \citenamefont {Ren}, \citenamefont {Luo}, \citenamefont {Wang},
			\citenamefont {Jin}, \citenamefont {Zhang},\ and\ \citenamefont
			{Cheng}}]{Song2013}%
		\BibitemOpen
		\bibfield  {author} {\bibinfo {author} {\bibfnamefont {N.N.}\ \bibnamefont
				{Song}}, \bibinfo {author} {\bibfnamefont {H.~T.}\ \bibnamefont {Yang}},
			\bibinfo {author} {\bibfnamefont {F.~Y.}\ \bibnamefont {Li}}, \bibinfo
			{author} {\bibfnamefont {Z.A.}\ \bibnamefont {Li}}, \bibinfo {author}
			{\bibfnamefont {W.}~\bibnamefont {Han}}, \bibinfo {author} {\bibfnamefont
				{X.}~\bibnamefont {Ren}}, \bibinfo {author} {\bibfnamefont {Y.}~\bibnamefont
				{Luo}}, \bibinfo {author} {\bibfnamefont {X.C.}\ \bibnamefont {Wang}},
			\bibinfo {author} {\bibfnamefont {C.~Q.}\ \bibnamefont {Jin}}, \bibinfo
			{author} {\bibfnamefont {X.Q.}\ \bibnamefont {Zhang}}, \ and\ \bibinfo
			{author} {\bibfnamefont {Z.H.}\ \bibnamefont {Cheng}},\ }\bibfield  {title}
		{\enquote {\bibinfo {title} {Interspacing dependence of spin-dependent
					variable range hopping for cold-pressed {Fe$_3$O$_4$} nanoparticles},}\
		}\href {\doibase http://aip.scitation.org/doi/full/10.1063/1.4804335}
		{\bibfield  {journal} {\bibinfo  {journal} {Journal of Applied Physics}\
			}\textbf {\bibinfo {volume} {113}},\ \bibinfo {pages} {184309} (\bibinfo
			{year} {2013})}\BibitemShut {NoStop}%
		\bibitem [{\citenamefont {Rani}\ \emph {et~al.}(2013)\citenamefont {Rani},
			\citenamefont {Kumar}, \citenamefont {Batoo},\ and\ \citenamefont
			{Singh}}]{Rani2013}%
		\BibitemOpen
		\bibfield  {author} {\bibinfo {author} {\bibfnamefont {R.}~\bibnamefont
				{Rani}}, \bibinfo {author} {\bibfnamefont {G.}~\bibnamefont {Kumar}},
			\bibinfo {author} {\bibfnamefont {K.~M.}\ \bibnamefont {Batoo}}, \ and\
			\bibinfo {author} {\bibfnamefont {M.}~\bibnamefont {Singh}},\ }\bibfield
		{title} {\enquote {\bibinfo {title} {Electric and dielectric study of zinc
					substituted cobalt nanoferrites prepared by solution combustion method},}\
		}\href {http://pubs.sciepub.com/ajn/1/1/3} {\bibfield  {journal} {\bibinfo
				{journal} {American Journal of Nanomaterials}\ }\textbf {\bibinfo {volume}
				{1}},\ \bibinfo {pages} {9--12} (\bibinfo {year} {2013})}\BibitemShut
		{NoStop}%
		\bibitem [{\citenamefont {Gul}\ \emph {et~al.}(2007)\citenamefont {Gul},
			\citenamefont {Abbasi}, \citenamefont {Amin}, \citenamefont {ur~Rehman},\
			and\ \citenamefont {Maqsood}}]{Gul2007}%
		\BibitemOpen
		\bibfield  {author} {\bibinfo {author} {\bibfnamefont {I.H.}\ \bibnamefont
				{Gul}}, \bibinfo {author} {\bibfnamefont {A.Z.}\ \bibnamefont {Abbasi}},
			\bibinfo {author} {\bibfnamefont {F.}~\bibnamefont {Amin}}, \bibinfo {author}
			{\bibfnamefont {M.~Anis}\ \bibnamefont {ur~Rehman}}, \ and\ \bibinfo {author}
			{\bibfnamefont {A.}~\bibnamefont {Maqsood}},\ }\bibfield  {title} {\enquote
			{\bibinfo {title} {Structural, magnetic and electrical properties of
					{Co$_{1-x}$Zn$_x$Fe$_2$O$_4$} synthesized by co-precipitation method},}\
		}\href {\doibase https://doi.org/10.1016/j.jmmm.2006.08.005} {\bibfield
			{journal} {\bibinfo  {journal} {Journal of Magnetism and Magnetic Materials}\
			}\textbf {\bibinfo {volume} {311}},\ \bibinfo {pages} {494 -- 499} (\bibinfo
			{year} {2007})}\BibitemShut {NoStop}%
		\bibitem [{\citenamefont {Reddy}\ \emph {et~al.}(1999)\citenamefont {Reddy},
			\citenamefont {Mohan}, \citenamefont {Boyanov},\ and\ \citenamefont
			{Ravinder}}]{Ramana1999}%
		\BibitemOpen
		\bibfield  {author} {\bibinfo {author} {\bibfnamefont {A.V.~Ramana}\
				\bibnamefont {Reddy}}, \bibinfo {author} {\bibfnamefont {G~Ranga}\
				\bibnamefont {Mohan}}, \bibinfo {author} {\bibfnamefont {B.S}\ \bibnamefont
				{Boyanov}}, \ and\ \bibinfo {author} {\bibfnamefont {D}~\bibnamefont
				{Ravinder}},\ }\bibfield  {title} {\enquote {\bibinfo {title} {Electrical
					transport properties of zinc-substituted cobalt ferrites},}\ }\href {\doibase
			https://doi.org/10.1016/S0167-577X(98)00234-1} {\bibfield  {journal}
			{\bibinfo  {journal} {Materials Letters}\ }\textbf {\bibinfo {volume} {39}},\
			\bibinfo {pages} {153 -- 165} (\bibinfo {year} {1999})}\BibitemShut {NoStop}%
		\bibitem [{\citenamefont {Barzilai}\ \emph {et~al.}(1981)\citenamefont
			{Barzilai}, \citenamefont {Goldstein}, \citenamefont {Balberg},\ and\
			\citenamefont {Helman}}]{Barzilai1981}%
		\BibitemOpen
		\bibfield  {author} {\bibinfo {author} {\bibfnamefont {S.}~\bibnamefont
				{Barzilai}}, \bibinfo {author} {\bibfnamefont {Y.}~\bibnamefont {Goldstein}},
			\bibinfo {author} {\bibfnamefont {I.}~\bibnamefont {Balberg}}, \ and\
			\bibinfo {author} {\bibfnamefont {J.~S.}\ \bibnamefont {Helman}},\ }\bibfield
		{title} {\enquote {\bibinfo {title} {{Magnetic and transport properties of
						granular cobalt films}},}\ }\href
		{http://prb.aps.org/abstract/PRB/v23/i4/p1809{\_}1} {\bibfield  {journal}
			{\bibinfo  {journal} {Physical Review B}\ }\textbf {\bibinfo {volume} {23}},\
			\bibinfo {pages} {1809} (\bibinfo {year} {1981})}\BibitemShut {NoStop}%
		\bibitem [{\citenamefont {Balcells}\ \emph {et~al.}(1998)\citenamefont
			{Balcells}, \citenamefont {Fontcuberta}, \citenamefont {Martinez},\ and\
			\citenamefont {Obradors}}]{Balcells1998}%
		\BibitemOpen
		\bibfield  {author} {\bibinfo {author} {\bibfnamefont {L.~L.}\ \bibnamefont
				{Balcells}}, \bibinfo {author} {\bibfnamefont {J.}~\bibnamefont
				{Fontcuberta}}, \bibinfo {author} {\bibfnamefont {B.}~\bibnamefont
				{Martinez}}, \ and\ \bibinfo {author} {\bibfnamefont {X.}~\bibnamefont
				{Obradors}},\ }\bibfield  {title} {\enquote {\bibinfo {title} {Magnetic
					surface effects and low-temperature magnetoresistance in manganese
					perovskites},}\ }\href {\doibase 10.1088/0953-8984/10/8/020} {\bibfield
			{journal} {\bibinfo  {journal} {Journal of Physics: Condensed Matter}\
			}\textbf {\bibinfo {volume} {10}},\ \bibinfo {pages} {1883} (\bibinfo {year}
			{1998})}\BibitemShut {NoStop}%
		\bibitem [{\citenamefont {Moodera}\ \emph {et~al.}(1997)\citenamefont
			{Moodera}, \citenamefont {Gallagher}, \citenamefont {Robinson},\ and\
			\citenamefont {Nowak}}]{Moodera1997}%
		\BibitemOpen
		\bibfield  {author} {\bibinfo {author} {\bibfnamefont {J.~S.}\ \bibnamefont
				{Moodera}}, \bibinfo {author} {\bibfnamefont {E.~F.}\ \bibnamefont
				{Gallagher}}, \bibinfo {author} {\bibfnamefont {K.}~\bibnamefont {Robinson}},
			\ and\ \bibinfo {author} {\bibfnamefont {J.}~\bibnamefont {Nowak}},\
		}\bibfield  {title} {\enquote {\bibinfo {title} {Optimum tunnel barrier in
					ferromagnetic-insulator-ferromagnetic tunneling structures},}\ }\href
		{\doibase 10.1063/1.118168} {\bibfield  {journal} {\bibinfo  {journal}
				{Applied Physics Letters}\ }\textbf {\bibinfo {volume} {70}},\ \bibinfo
			{pages} {3050--3052} (\bibinfo {year} {1997})}\BibitemShut {NoStop}%
		\bibitem [{\citenamefont {Joo}\ \emph {et~al.}(2014)\citenamefont {Joo},
			\citenamefont {Jung}, \citenamefont {Jun}, \citenamefont {Kim}, \citenamefont
			{Shin}, \citenamefont {Hong}, \citenamefont {Lee},\ and\ \citenamefont
			{Rhie}}]{Joo2014}%
		\BibitemOpen
		\bibfield  {author} {\bibinfo {author} {\bibfnamefont {S.}~\bibnamefont
				{Joo}}, \bibinfo {author} {\bibfnamefont {K.~Y.}\ \bibnamefont {Jung}},
			\bibinfo {author} {\bibfnamefont {K.~I.}\ \bibnamefont {Jun}}, \bibinfo
			{author} {\bibfnamefont {D.~S.}\ \bibnamefont {Kim}}, \bibinfo {author}
			{\bibfnamefont {K.~H.}\ \bibnamefont {Shin}}, \bibinfo {author}
			{\bibfnamefont {J.~K.}\ \bibnamefont {Hong}}, \bibinfo {author}
			{\bibfnamefont {B.~C.}\ \bibnamefont {Lee}}, \ and\ \bibinfo {author}
			{\bibfnamefont {K.}~\bibnamefont {Rhie}},\ }\bibfield  {title} {\enquote
			{\bibinfo {title} {Spin-filtering affect of thin {Al$_2$O$_3$} barrier on
					tunneling magnetoresistance},}\ }\href {\doibase 10.1063/1.4870812}
		{\bibfield  {journal} {\bibinfo  {journal} {Applied Physics Letters}\
			}\textbf {\bibinfo {volume} {104}},\ \bibinfo {pages} {152407} (\bibinfo
			{year} {2014})}\BibitemShut {NoStop}%
	\end{thebibliography}


\end{document}